\documentclass[useAMS,usenatbib,fleqn]{mn2e}

\usepackage{aas_macros}

\usepackage{epsfig}
\usepackage{epstopdf}
\usepackage{subfigure}
\usepackage{amsmath}    % need for subequations
\usepackage{longtable}
\usepackage{dcolumn}
\usepackage{amssymb}

\usepackage{upgreek} % my addition - for displaying upright pi (\uppi)

\usepackage{wrapfig}

\DeclareGraphicsExtensions{.jpg,.eps,.EPS,.png,.pdf,.ps}

\usepackage{colordvi}
\usepackage{ulem}

\RequirePackage[colorlinks=true
,urlcolor=blue
,anchorcolor=blue
,citecolor=blue
,filecolor=blue
,linkcolor=blue
,menucolor=blue
,pagecolor=blue
,linktocpage=true
,pdfproducer=medialab
]{hyperref}

\defcitealias{Margalit&Metzger16}{MM16}

\newcommand       \be          {\begin{eqnarray}}
\newcommand       \ee          {\end{eqnarray}}

\usepackage[usenames,dvipsnames]{xcolor}

\title[WD-NS Mergers to Pulsar Planets]{Merger of a White Dwarf-Neutron Star Binary to $10^{29}$ Carat Diamonds: Origin of the Pulsar Planets}

\author[B. Margalit \& B.~D. Metzger]
{Ben Margalit\thanks{E-mail: \href{mailto:btm2134@columbia.edu}{btm2134@columbia.edu}} and Brian D. Metzger \\ 
\normalsize Columbia Astrophysics Laboratory, Columbia University, 538 West 120th St., New York, NY 10027
}

\date{}

\begin{document}
\maketitle

\begin{abstract}

We show that the merger and tidal disruption of a C/O white dwarf (WD) by a neutron star (NS) binary companion provides a natural formation scenario for the PSR B1257+12 planetary system.  Starting with initial conditions for the debris disk produced of the disrupted WD, we model its long term viscous evolution, including for the first time the effects of mass and angular momentum loss during the early radiatively inefficient accretion flow (RIAF) phase and accounting for the unusual C/O composition on the disk opacity.  For plausible values of the disk viscosity $\alpha \sim 10^{-3}-10^{-2}$ and the RIAF mass loss efficiency, we find that the disk mass remaining near the planet formation radius at the time of solid condensation is sufficient to explain the pulsar planets.  Rapid rocky planet formation via gravitational instability of the solid carbon-dominated disk is facilitated by the suppression of vertical shear instabilities due to the high solid-to-gas ratio.  Additional evidence supporting a WD-NS merger scenario includes (1) the low observed occurrence rate of pulsar planets ($\lesssim 1\%$ of NS birth), comparable to the expected WD-NS merger rate; (2) accretion by the NS during the RIAF phase is sufficient to spin PSR B1257+12 up to its observed 6 ms period; (3) similar models of `low angular momentum' disks, such as those produced from supernova fallback, find insufficient mass reaching the planet formation radius.  The unusually high space velocity of PSR B1257+12 of $\gtrsim 326\,{\rm km\,s}^{-1}$ suggests a possible connection to the Calcium-rich transients, dim supernovae which occur in the outskirts of their host galaxies and were proposed to result from mergers of WD-NS binaries receiving SN kicks.  The C/O disk composition implied by our model likely results in carbon-rich planets with diamond interiors.
\end{abstract}


\section{Introduction}
The millisecond pulsar PSR B1257+12 is famous for hosting the first known extra-solar planets, as discovered through its timing residuals \citep{Wolszczan&Frail92}. Continuing observations have confirmed the existence of two earth-mass exoplanets and one moon size body orbiting PSR B1257+12 with masses $3.9 M_\oplus$, $4.3 M_\oplus$, $0.02 M_\oplus$ and semi-major axes $0.46\,{\rm AU}$, $0.36\,{\rm AU}$, $0.19\,{\rm AU}$, respectively \citep{Wolszczan94,Konacki&Wolszczan03}. The nearly coplanar orbits of the earth-mass planets, along with their $\sim$3:2 resonance and small eccentricity ($e \simeq 0.02$), provide strong evidence for a disk formation scenario \citep{Konacki&Wolszczan03}. Although a number of general formation scenarios have been proposed \citep[][and references therein]{Phinney&Hansen93,Podsiadlowski93}, the origin of the pulsar planets remains a mystery.

The fundamental reason that the PSR B1257+12 planetary system poses such a theoretical challenge, and is inherently different from our own solar system or the multitude of exoplanetary systems discovered since, is the fact that the central body is a neutron star (NS).  Standard planet formation theory focuses on bodies formed in the gaseous disk that exists concurrent with the birth of the stellar system (e.g., the progenitor of the NS).  However, these theories encounter severe difficulties in explaining the PSR B1257+12 system because it is unclear how the observed planets could have survived the supernova (SN) explosion responsible for the NS, or the preceding giant phase during which the planets would have been engulfed in an extended stellar envelope (e.g.~\citealt{Nordhaus+10}).  It is thus conventionally believed that the PSR B1257+12 planets must have formed after the SN, thereby requiring a novel formation mechanism.

\cite{Podsiadlowski93} overviews the range of proposed origins for the pulsar planets.
Beyond elucidating the planet formation mechanism itself, a successful model should explain why PSR B1257+12 is a recycled millisecond pulsar (\citealt{Miller&Hamilton01}), as well as the dearth of planets around the vast majority of other pulsars (\citealt{Kerr+15}; Fig.~\ref{fig:KerrConstraints}).  Two leading explanations for the gaseous disk out of which the planets form are: (a) the fall-back accretion of bound stellar debris following the SN explosion (e.g.~\citealt{Menou+01}), and (b) the tidal disruption of a stellar object into a gaseous disk following a close encounter with the NS.  Variants of the latter include the disruption of a white dwarf (WD) by a binary NS companion; a WD-WD merger in which one WD is disrupted and the other gains mass and collapses to a NS; and the collision of the NS with a non-degenerate star, such as a binary companion (following the SN birth kick received by the NS). 

\cite{Phinney&Hansen93} explore the viability of a wide range of such models, using a general analytic framework to describe the long-term viscous spreading of the comparatively promptly formed gaseous disk (\citealt{Pringle81}). They found that both SN fallback and disruption models are at least marginally capable of placing sufficient mass on the correct radial scale to explain the PSR B1257+12 system. \cite{Currie&Hansen07} built on this work by including additional physical ingredients, such as irradiation and layered accretion, and by numerically solving for the disk evolution. 
These authors found that tidal disruption models typically underproduce the amount of solids required at $\sim 1~{\rm AU}$ compared with SN fallback models, in large part due to the assumption of solar metallicity in the tidal disruption models as compared to the metal-rich fall back case.  In followup work, \cite{Hansen+09} study the planet assembly process from smaller bodies, using initial conditions motivated by the disk models of \cite{Currie&Hansen07}.

In this work we explore the merger of WD-NS binaries as a formation channel for the PSR B1257+12 planetary system.   We describe the long-term evolution of the remnant gaseous disks from such mergers using an approximate analytic approach, which extends the work of \cite{Phinney&Hansen93}, however incorporating several physical aspects unique to this model.
In particular, we impose initial conditions informed by our recent study of the early phase of WD-NS disk evolution \citep[hereafter \citetalias{Margalit&Metzger16}]{Margalit&Metzger16}. We also model the previously ignored radiatively-inefficient accretion flow (RIAF) phase at early times, which leads to an important mass sink in the form of disk outflows.  We also account for the unique C/O composition predicted by this scenario.

The early phases of a WD-NS merger have been previously studied as sources of gamma-ray bursts \citep{Fryer+99,King+07,Paschalidis+11} and SN-like optical transients (\citealt{Metzger12}; \citealt{Fernandez&Metzger13}; \citetalias{Margalit&Metzger16}). The stages in the evolution of such systems are briefly summarized as follows (e.g. \citetalias{Margalit&Metzger16}).  A detached WD-NS binary slowly inspirals due to gravitational wave emission, before overflowing its Roche lobe at a separation of $\sim 10^9~{\rm cm}$.  If the ratio of the WD to the NS mass is sufficiently large (e.g., $\gtrsim 0.2-0.5$; \citealt{Bobrick+16}), then mass transfer from the WD is unstable, and the ensuing phase of runaway accretion results in the WD being tidally disrupted \citep{Verbunt&Rappaport88} into a thick and dense torus surrounding the NS (e.g.~\citealt{Fryer+99}).  During the earliest stages of disk evolution ($\sim$ minutes-hours), dynamically important nuclear burning occurs in the disk midplane \citep{Metzger12}.  Outflows from the disk reduce the mass reaching the central NS and provide a necessary source of cooling, which offsets gravitational (viscous) and nuclear heating.

As a pulsar planet formation scenario, a WD-NS merger is appealing for several reasons.  First, it naturally explains the millisecond rotation period of PSR B1257+12, as the initially old NS is spun-up due to the accretion of debris from the disrupted WD (\citealt{vandenHeuvel&Bonsema84}; \citealt{Ruderman&Shaham85}; \citetalias{Margalit&Metzger16}).  Second, the rates of WD-NS mergers (a fraction $\sim 10^{-3}-10^{-2}$ of the core collapse SN rate; \citealt{OShaughnessy&Kim10,Kim+15,Bobrick+16}) agree well with the observed paucity of pulsar planet systems. 
\cite{Kerr+15} recently surveyed a sample of 151 young pulsars and found no evidence for additional planetary systems, implying that only a small fraction $\lesssim 10^{-2}$ of pulsars host planets.  This agreement contrasts with SN fallback models, which at least naively would predict pulsar planetary systems to be common.  Finally, the high metallicity of disks formed from WD debris can lead to a large fraction of the disk mass forming solid rocks, circumventing efficiency problems associated with solar metallicity models.  One intriguing consequence of the predicted carbon-rich composition of the pulsar planets is that they are in essence enormous ($\sim 10^{29}$ Carat) diamonds (e.g., \citealt{Kuchner&Seager05}).

This paper is organized as follows.  We begin by outlining our viscous disk model in \S~\ref{sec:DiskModel}. In the following section we present detailed results for a sample model, and investigate whether disk conditions are conducive to planet-formation by conducting a parameter-space survey (\S~\ref{sec:DiskResults}). We continue by discussing a possible planet-formation mechanism specific to our WD-NS merger scenario (\S~\ref{sec:PlanetFormation}). Finally, in \S~\ref{sec:Discussion} we conclude and discuss possible implications and observational signatures of our model.

\section{Disk Evolution Model} \label{sec:DiskModel}

The most basic feature of accretion disk evolution follows from angular momentum conservation --- globally, the disk must spread as it loses mass in order to conserve angular momentum (\citealt{Pringle81}).  If the angular momentum carried away by the lost mass is negligible (which is {\it not} true during the early RIAF phase following a WD-NS merger), then the mass-averaged disk radius $R_{\rm d}$ increases with decreasing total mass $M_{\rm d}$ as
\begin{equation} \label{eq:RdMd_radiative}
R_{\rm d} \propto M_{\rm d}^{-2} .
\end{equation}
This process of `viscous spreading' allows planetary bodies to form on $\sim$ AU radial scales from an initial disk of size $\sim 10^{9}$ cm $\approx 10^{-4}$ AU.

In greater detail, the disk surface density $\Sigma$ evolves according to a diffusion equation \citep{Pringle81}
\begin{equation}
\partial_t \Sigma = \frac{3}{r} \partial_r \left[ \nu \frac{\partial \ln \left( \nu r^2 \Sigma \Omega \right)}{\partial \ln r} \right] - \dot{\Sigma}_{\rm w} ,
\end{equation}
where $\dot{\Sigma}_{\rm w}$ is a sink term accounting for local mass loss in the form of disk outflows.
Outwards angular momentum transfer is commonly described by an effective kinematic viscosity 
\begin{equation}
\nu = \alpha c_{\rm s} H \approx \alpha c_{\rm s}^{2}/\Omega = \alpha \theta^{2}r v_k,
\end{equation}
where $c_{\rm s} \equiv \sqrt{P/\rho}$ is the isothermal sound speed in the disk midplane, $H = c_{\rm s}/\Omega$ the disk scale-height, $\Omega \approx \Omega_{\rm k} = v_k/r = \sqrt{G M_{\rm NS} / r^3}$ is the orbital angular velocity, and 
\begin{equation}
\theta \equiv {H}/{r} = {c_{\rm s}}/{v_{\rm k}},
\end{equation}
is the disk aspect ratio.  The dimensionless parameter $\alpha$ parameterizes the strength of the viscosity \citep{Shakura&Sunyaev73}.  When the disk is ionized, turbulence due to the magnetorotational instability (MRI) provides a physical mechanism driving angular momentum transport \citep[][]{Balbus&Hawley91}, with values of $\alpha \gtrsim 0.01-0.1$ motivated by simulations of the MRI \citep[e.g.][]{Davis+10} and observations (\citealt{King+07}).  Once the midplane temperature decreases to $\lesssim 1000$ K and the disk material recombines to become neutral, the MRI may be suppressed or confined to a thin `active' ionized layer on the disk surface (\citealt{Gammie96}, \citealt{Hansen02}).  However, the high intensity of cosmic rays and ionizing radiation in the vicinity of an accreting neutron star will substantially reduce the dead zone as compared to the standard proto-stellar case.  Turbulent eddies in the active region may overshoot the active/dead boundary and transport mass also within the dead zone (e.g., \citealt{Fleming&Stone03}).

In thermal equilibrium, a local balance exists between the heating and cooling rates per unit surface area of the disk, 
$\dot{q}^+ = \dot{q}^-$.  During the early phases of disk evolution, viscous dissipation provides the dominant heat source,
\begin{equation} \label{eq:qdot_visc}
\dot{q}^+_{\rm visc} = \nu \Sigma \Omega^2 \left(\frac{\partial \ln \Omega}{\partial \ln r} \right)^2 
\approx \frac{9}{4} \alpha \theta^2 \Sigma \Omega_{\rm k}^3 r^2,
\end{equation}
while at later times irradiation of the outer disk from the central accretion flow dominates.  The competing cooling rate $\dot{q}^-$ is set by disk outflows/radial advection at early times, and radiative cooling at late times.  

Specifying an equation of state (EOS) closes the equations, allowing a full solution of the disk radial structure and temporal evolution in terms of $\Sigma$, $\theta$, and the midplane temperature $T$.
Two limiting cases, corresponding to the midplane pressure being dominated by gas or radiation, respectively, give temperatures of
\begin{equation}
T = 
\begin{cases}
(\mu m_{\rm p} / k_{\rm B}) \theta^2 \Omega^2 r^2 ~; ~~& \mathrm{gas} \\
\left[ (3 / 2 a) \theta \Omega^2 r \Sigma \right]^{1/4} ~; & \mathrm{radiation} ,
\end{cases}
\end{equation}
where $\mu$ is the mean molecular weight.  In analytic estimate we adopt a value of $\mu = 13$ corresponding to neutral gas of half carbon and half oxygen composition, although $\mu$ can be lower at early times (high temperatures) when the gas is ionized or higher at later times (lower temperature) after molecular CO forms.  

\subsection{Evolution Stages} \label{subsec:EvolutionStages}
\label{sec:stages}

\begin{figure} 
\centering
\epsfig{file=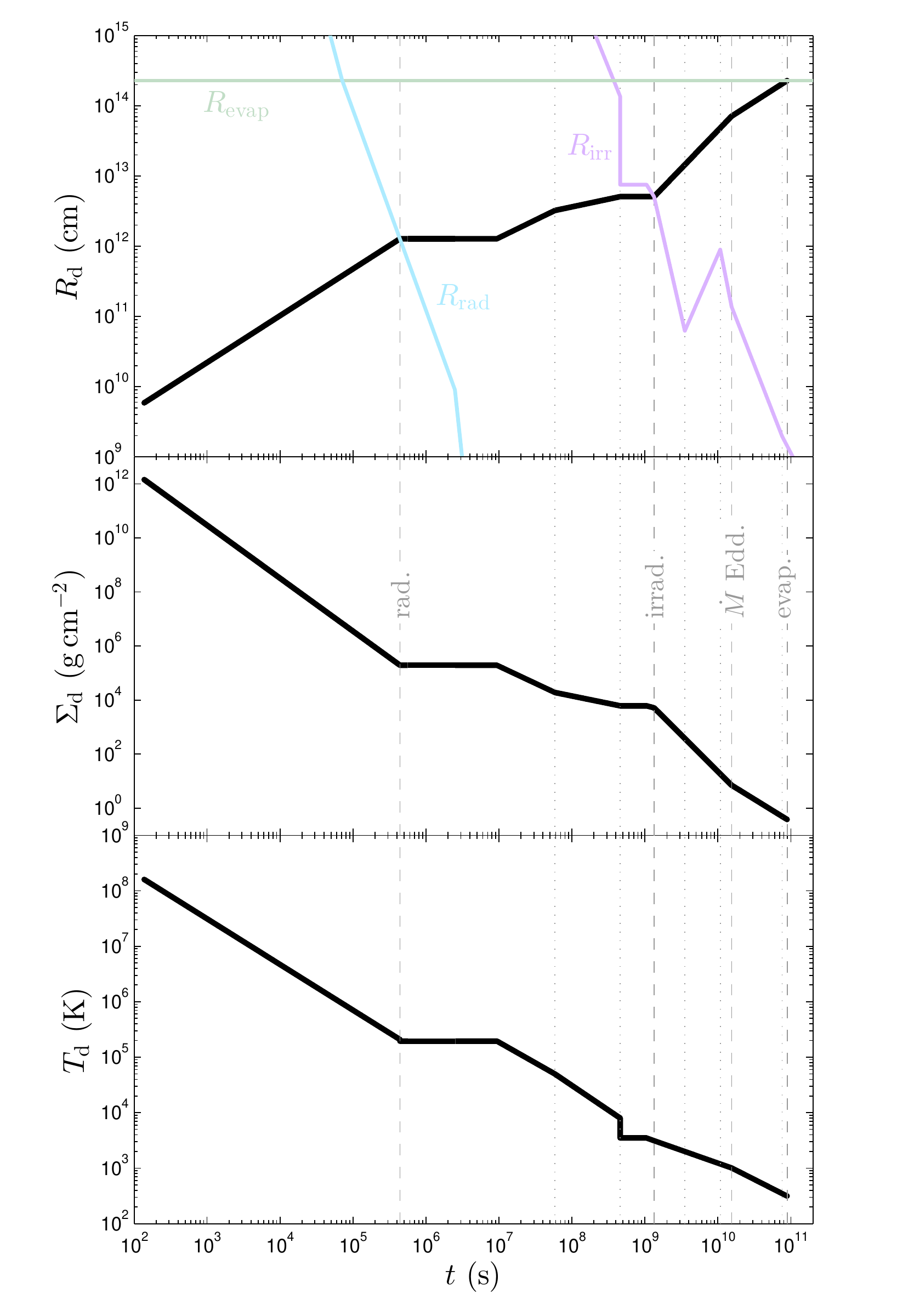,angle=0,width=0.5\textwidth}
\caption{Temporal evolution of the remnant accretion disk from a WD-NS merger, corresponding to the fiducial model of a 0.6$M_{\odot}$ C/O WD for a viscosity $\alpha = 0.1$, with initial conditions from \citetalias{Margalit&Metzger16}.  Panels show the outer disk radius (top), surface density (middle), and temperature (panel).  Vertical dashed lines indicate important transitions in the disk accretion regime, whereas vertical dotted lines show transitions in the opacity law (which only affect the evolution during the viscously heated radiative phase). The top panel also shows important transition radii $R_{\rm rad}$, $R_{\rm irr}$, and $R_{\rm evap}$ (see description in $\S\ref{sec:stages}$).} \label{fig:Evolution}
\end{figure}

The dominant sources of heating and cooling change as the disk viscously spreads outwards.  Here, we overview the evolutionary stages of the disk, before describing each in greater detail (see Fig.~\ref{fig:Evolution} for an example solution presented later).  

\begin{enumerate}
\item{Immediately after its formation, the dense torus is unable to cool effectively via radiation because the photon diffusion timescale out of the disk midplane is much longer than the viscous timescale.  During this `RIAF' phase, heating due to viscous dissipation and nuclear burning are offset by cooling due to radial advection and by launching powerful disk outflows, which carry away both energy and mass.}

\item{The disk evolves in the RIAF regime until the midplane density decreases to the point that radiation is no longer trapped on the accretion timescale.  After this point, radiative cooling takes over and the disk transitions to a standard thin configuration.  In this `Viscously-Heated Radiative' phase, the disk cooling rate depends on its optical depth and hence on the opacity law $\bar{\kappa}(\rho,T)$.  The resulting evolution is rich, traversing a diversity of phases as the decreasing midplane temperature/density evolution carry it through different opacity regimes.}

\item{As the disk continues to evolve and the accretion rate decreases, heating due to irradiation from the innermost accretion flow eventually comes to exceed internal viscous heating.  Once the disk enters this `Irradiated' phase, it becomes vertically isothermal and the midplane temperature no longer depends on $\bar{\kappa}$.  The disk evolution no longer depends on the opacity law, but a transition in its evolution can still occur when the accretion rate drops below the Eddington rate.  }

\item{The disk continues in the irradiated phase until finally dispersing at late times due to solid condensation and photo-evaporation.  Evaporation begins in full once the disk spreads to a sufficiently large radius $R_{\rm evap} \sim 10$ AU, where the sound speed of irradiated surface layers of the disk exceeds about ten percent of the escape speed.  In this `Photo-evaporation' phase, the disk expansion stalls at $R_{\rm evap}$ and its gas content drains exponentially on the viscous timescale, in contrast to the power-law evolution which characterizes the earlier accretion stages.}

\end{enumerate}

\subsubsection{RIAF Phase}

The accretion torus formed by the tidal disruption of the WD is initially extremely hot and sufficiently dense as to be unable to cool through radiative diffusion.  The early stages of this torus evolution were explored by \citetalias{Margalit&Metzger16}, who adopt an ADIOS type model \citep{Blandford&Begelman99} which presumes that a powerful wind of hot matter is blown off the disk.  The mass loss rate and wind kinetic energy at each radius are determined by the requirement that the winds carry away sufficient energy to locally  balance viscous and nuclear heating, such that the Bernoulli parameter of the disk midplane is regulated to a fixed value $\lesssim 0$ (bound disk). 

The radial dependence of the mass inflow rate during the RIAF can be approximated as a power-law,
\begin{equation}
\dot{M}_{\rm in} = \dot{M}_{\rm d} \left( \frac{r}{R_{\rm d}} \right)^p ,
\end{equation}
where hereafter all quantities with subscripts $X_{\rm d}$ are evaluated at the outer (characteristic) disk radius.  The exponent $0 \leq p < 1$ characterizes the disk mass loss, since continuity implies that the mass loss rate obey $\dot{M}_{\rm out} \propto \left[1-\left(R_{\rm NS}/R_{\rm d}\right)^p\right]$.
Hydrodynamical $\alpha$-disk and global MHD simulations of RIAFs typically find $p \gtrsim 0.5$ \citep{Stone+99,Igumenshchev+00,Hawley+01,Narayan+12,McKinney+12,Yuan+12}.

Importantly, the total angular momentum of the disk is not conserved during the RIAF phase due to the presence of disk winds.  Under the assumption that disk winds exert zero net torque on the disk (i.e. mass is lost with the same specific angular momentum as that of the disk midplane from the wind launching point), \citet{Metzger+08} show that
\begin{equation} \label{eq:RdMd_RIAF}
R_{\rm d} \propto M_{\rm d}^{-2/(2p+1)},
\end{equation}
which reduces to equation (\ref{eq:RdMd_radiative}) in the zero wind mass loss limit ($p = 0$).

As initial conditions for the present study, we make use of the `final' disk configurations of \citetalias{Margalit&Metzger16}, calculated out to several times the initial viscous timescale of the torus formed from the merger.  We extrapolate these `final' configurations to much later times of interest here, using the RIAF self-similar evolution described by \cite{Metzger+08} and summarized in Appendix~\ref{sec:Appendix_SelfSimilar}.  Except at very early times, the disk is dominated by radiation pressure during the RIAF phase, and becomes increasingly so with time.

\subsubsection{Viscously-Heated, Radiatively Cooled Phase}

Once the disk enters the radiative phase, the midplane cools via radiation diffusion at a rate given by
\begin{equation} \label{eq:qdot_rad}
\dot{q}^-_{\rm rad} = \frac{4 \sigma T^4}{3 f(\tau)} ,
\end{equation}
where 
\begin{equation} \label{eq:f_of_tau}
f(\tau) \approx \tau + {2}\tau^{-1}/{3} + {4}/{3} \approx 
\begin{cases} 
\tau; &\tau \geq \sqrt{2/3} \\ 
2 / 3 \tau; &\tau < \sqrt{2/3} ~.
\end{cases}
\end{equation}
is a function of the vertical optical depth ($\tau = \bar{\kappa} \Sigma / 2$), which results from solving the vertical radiative transfer problem \citep[e.g.][]{Sirko&Goodman03}, and $\bar{\kappa}$ is the Rosseland mean opacity.  Moving from high to low temperatures, relevant opacities include electron scattering, free-free/bound-free opacities of C/O-rich matter, and graphite dust.  Appendix~\ref{sec:Appendix_Opacity} describes the full opacity curve $\bar{\kappa}(\rho,T)$ implemented in our analysis, which we approximate in our analytic estimates by a series of broken power-laws of the form $\bar{\kappa} \propto \rho^l T^{-k}$ (the exponents $l$, $k$ enter in the self-similar disk solutions described in Appendix~\ref{sec:Appendix_SelfSimilar}).

During the RIAF phase, viscous heating is offset by radial advection and wind cooling.  However, as the disk evolves and its optical depth decreases, radiative losses eventually become dominant, and the disk becomes geometrically thin ($\theta_d \ll 1$).  This transition to the radiative phase occurs once\footnote{Note that the photon diffusion timescale is shorter than the viscous inflow timescale when this criterion is satisfied, justifying treating the radiative energy losses as truly being local.}  $\dot{q}^{-}_{\rm rad} = \eta_{\rm rad} \dot{q}^{+}_{\rm visc}$ with $\eta_{\rm rad} \lesssim 1$.  In the optically-thick, radiation-pressure dominated regime appropriate for the RIAF phase, this condition is satisfied once the disk radius exceeds the critical value
\begin{equation} \label{eq:R_rad}
R_{\rm rad} \approx \left(\frac{3}{2}\right)^4  \eta_{\rm rad}^2 \alpha^2 \theta^2 \tau^2 \frac{G M_{\rm NS}}{c^2} .
\end{equation}
Initially, the value of $R_{\rm rad}$ greatly exceeds the actual outer radius of the disk, $R_{\rm d}$.  However, as $R_{\rm rad}$ decreases and $R_{\rm d}$ grows, the outer disk transitions from RIAF to radiative once $R_{\rm d} \approx R_{\rm rad}$ (Fig.~\ref{fig:Evolution}).

The radiative transition would be continuous, were it not for the fact that the radiation pressure-dominated, radiative disk solution is thought to be unstable \citep{Lightman&Eardley74}.  Assuming this instability manifests, then following the radiative transition, the disk immediately (on a thermal timescale) collapses to the corresponding gas-pressure dominated solution satisfying $\dot{q}^-_{\rm rad}(T,\theta) = \dot{q}^+_{\rm visc}(\theta)$ for the same $\Sigma$, $R_{\rm d}$ \citep[see][for a similar discussion in the context of tidal disruption events]{Shen&Matzner14}.  Recent work by \citet{Jiang+16} suggests that opacity due to line transitions of iron may help stabilize the disk at temperatures $T \sim 2\times 10^{5}$ K close to those which characterize the radiative transition in our models; however, additional theoretical work is required to determine whether the disk can truly remain stable for the tens or hundreds of thermal times relevant to the viscous timescale evolution.

Another important discontinuity occurs if the disk cools to $T \sim 10^4~{\rm K}$ while still in the optically-thick, viscously-heated radiative phase. At such temperatures the disk is susceptible to the recombination instability, induced by the sudden drop in opacity once the gas begins to recombine (see Fig. \ref{fig:OpacityCurve}).  Modeling the opacity during this transition as a steep power-law, $\bar{\kappa} \propto T^{k}$, one can show that (in the optically thick regime) $\partial \ln \dot{q}^+_{\rm visc} / \partial \ln T > \partial \ln \dot{q}^-_{\rm rad} / \partial \ln T$, implying that the disk is thermally unstable \citep[][]{Piran78}.

The recombination instability has been invoked as a possible origin of the limit-cycle behavior observed in dwarf novae (e.g.~\citealt{Cannizzo+88, Coleman+16}). Indeed, as long as some fixed (and sufficiently large) accretion rate $\dot{M}_{\rm d}$ is enforced as an external boundary condition, the instability region will exhibit cyclical behavior, oscillating between the ionized and neutral solutions. If however, as in our case, the instability occurs first at the outer disk, and external mass is not supplied to the system (i.e. $\dot{M}_{\rm d}$ is not externally forced), then the result is a single transition between the ionized and neutral solutions.

\subsubsection{Irradiation-Heated, Radiatively Cooled Phase}

Irradiated disks have been widely studied in the case of protoplanetary disks illuminated by the central star or a surrounding cluster (e.g.~\citealt{Chiang&Goldreich97}). In our scenario, the disk is illuminated primarily by the accretion luminosity from the inner disk onto the central NS. 

Assuming that the inner disk is unobscured from the viewpoint of the outer disk, the heating rate per area due to irradiation is approximately given by
\begin{align} \label{eq:qdot_irr}
\dot{q}^+_{\rm irr} 
= \theta \left(\frac{\partial \ln \theta}{\partial \ln r}\right) \frac{G M_{\rm NS} }{4 \pi r^2 R_{\rm NS}} \min\left( \dot{M}, \dot{M}_{\rm Edd}\right) ~,
\end{align}
where the factors including $\theta$ account for the geometric cross-section illuminated by the central source, and the minimum accounts for the fact that the NS cannot radiate above the Eddington limit, $\dot{M}_{\rm Edd} \simeq 2 \times 10^{18} \left(R_{\rm NS}/10^6\,{\rm cm}\right) \left(\bar{\kappa}/0.2\,{\rm cm}^2\,{\rm g}^{-1}\right)^{-1} {\rm g}~{\rm s}^{-1}$.

In practice, when the disk is accreting at highly super-Eddington rates, the accretion luminosity from the inner disk could be channeled along the rotation axis by the thick inner torus (e.g.~\citealt{Jiang+14}, \citealt{Sadowski&Narayan15}), away from the outer disk.   There is also some evidence in AGN that the outer disk may be shielded from X-rays emitted by the inner disk, even at sub-Eddington accretion rates (e.g.~\citealt{Luo+15}).  To account for the possibility of shielding of the inner disk during the super-Eddington phase, we also consider models in which we set $\dot{q}^+_{\rm irr}  = 0$ for $\dot{M} \gtrsim \dot{M}_{\rm Edd}$. 

Solving the radiative transfer equation for an externally illuminated disk illustrates that the disk develops a nearly isothermal vertical profile, in contrast to when the disk is heated from the midplane (e.g.~\citealt{Kratter+10}).  As the radiative cooling rate no longer depends on $\tau$ (i.e.~$f(\tau) \to 1$ in equation~\ref{eq:qdot_rad}), irradiated disk solutions do not depend on the opacity.  However, from equation (\ref{eq:qdot_irr}) note that the time evolution does depend on whether $\dot{M}$ is larger or smaller than $\dot{M}_{\rm Edd}$.

The transition from viscous- to irradiation-dominated occurs when $\dot{q}^+_{\rm irr} = \dot{q}^+_{\rm visc} f(\tau)$, once the disk spreads to a characteristic radius
\begin{align}
R_{\rm irr} &= \frac{3}{2} R_{\rm NS} f(\tau) \max \left( 1, \frac{\dot{M}}{\dot{M}_{\rm Edd}} \right) \theta^{-1}  \\ \nonumber
&\times \left(\frac{\partial \ln \theta}{\partial \ln r}\right)^{-1} \left[ 1 + \left(\frac{\partial \ln \Sigma}{\partial \ln r}\right) + 2\left(\frac{\partial \ln \theta}{\partial \ln r}\right) \right]^{-1},
\end{align}
where we have used equation (\ref{eq:qdot_irr}).  Again, although initially $R_{\rm irr}$ greatly exceeds the physical extent of the disk $R_{\rm d}$, its value quickly decreases as the disk optical depth and accretion rate drop, such that the transition to the irradiated regime occurs at $R_{\rm d} = R_{\rm irr}$ (Fig.~\ref{fig:Evolution}).  If irradiation of the outer disk is blocked during the super-Eddington phase, then the transition instead coincides with the sub-Eddington transition.  

\subsubsection{Photo-Evaporation Phase}

The disk begins to evaporate due to photo-ionization heating once it spreads to radii approaching the so-called gravitational radius (\citealt{Hollenbach+94,Adams+04}),
\begin{equation}
R_{\rm g} = \frac{G M_{\rm NS}}{c_{\rm s, g}^2} \approx 400\,{\rm AU},
\end{equation}
at which the escape velocity $\sim (GM_{\rm NS}/r)^{1/2}$ equals the sound speed $c_{\rm s, g} \approx 2$ km s$^{-1}$ of photo-ionized gas with an approximately fixed temperature $\approx 5\times 10^{3}$ K set by the balance between photo-ionization and line cooling (e.g.~\citealt{Melis+10}, for a metal-enriched composition). 

Naively, one would expect photo-evaporation to become relevant only once the disk spreads to $R_{\rm d} \sim R_{\rm g}$.  However, more detailed estimates consider the exponential profile of the disk atmosphere and the fact that outflowing matter is accelerated further after detaching from the disk, as it is subjected to further external irradiation.  \cite{Adams+04} find that the mass evaporation rate from the outer disk edge can be expressed as (for $R_{\rm d} < R_{\rm g}$)
\begin{equation}
\dot{M}_{\rm evap} \approx 2 \pi R_{\rm d} \Sigma_{\rm d} c_{\rm s,g} \left(\frac{R_{\rm g}}{R_{\rm d}}\right)^2 e^{-R_{\rm g} / 2 R_{\rm d}} ~.
\end{equation}

The sensitive dependence of $\dot{M}_{\rm evap}$ on $R_{\rm d} / R_{\rm g}$ shows that, as the disk expands approaching $R_{\rm g}$, the mass loss rate rises exponentially.  Conversely, the evaporation timescale, $t_{\rm evap} \sim M_{\rm d} / \dot{M}_{\rm evap}$, decreases exponentially. This elucidates the subsequent evolution --- once the disk expands to the critical radius $R_{\rm evap}$ at which the evaporation timescale equals the accretion timescale, the disk evolution stalls.

Up to inessential numerical factors of order unity, the evaporation disk radius $R_{\rm evap}$ at which $t_{\rm evap} = t_{\rm visc}$ is determined by the solution of the transcendental equation
\begin{equation}
x^{-3/2} e^{x/2} = 1 / \alpha \theta_{\rm d}^2 ~,
\end{equation}
which in the range $\alpha \theta_{\rm d}^2 \sim 10^{-10}-10^{-2}$ is approximately given by
\begin{equation}
{R_{\rm g}}/{R_{\rm evap}} \equiv x \simeq -5.024 \log_{10} \left( \alpha \theta_{\rm d}^2 \right) + 8.283.
\end{equation}
For typical parameters, this yields values of $R_{\rm evap} \sim $ of a few percent of $R_{\rm g}$, corresponding to tens of AU.

Since the disk radius $R_{\rm d} \approx R_{\rm evap}$ is regulated to an approximately fixed value in the photo-evaporation phase, the viscous timescale (which depends most sensitively on radius) also remains approximately constant.  This breaks the self-similarity of the disk solution, resulting in the exponential evaporation/accretion of the remaining disk mass over the viscous timescale 
\begin{align}
t_{\rm evap} &\sim \left.\frac{M_{\rm d}}{\dot{M}_{\rm d}}\right|_{R_{\rm evap}} \approx \left.\frac{7A \mu m_{\rm p} \Omega r^{2}}{9 \alpha k_{\rm B} T}\right|_{R_{\rm evap}}  \nonumber \\
&\simeq 4\,{\rm Myr} \left(\frac{\alpha}{0.01}\right)^{-1}\left(\frac{R_{\rm evap}}{\rm 10 AU}\right)^{1/2}\left(\frac{\left.T\right|_{R_{\rm evap}}}{100\,{\rm K}}\right)^{-1}
\end{align} 
at the time of the transition to the evaporative phase, where we have taken a characteristic value of $T \approx 100$ K as an estimate of the temperature at the evaporative transition (see equation~\ref{eq:Tp_irr}), and $A \approx 3$ is a constant relating $M_{\rm d}$ to the local mass at $R_{\rm d}$ (equation \ref{eq:A}).

We neglect additional mass sinks due to solid condensation on the disk evolution, as was included, e.g., in \citet{Currie&Hansen07}.  Although solid formation is obviously relevant to planet formation, neglecting its effect on the disk evolution should not appreciably alter our results, in part because roughly half of the disk mass is comprised of oxygen and must remain in gaseous form (a significant fraction may also remain trapped in inert CO).  There is also some evidence that the solid condensation responsible for forming planets in the PSR B1257+12 system happened quickly once the appropriate conditions were first reached near the outer radius of the disk ($\S\ref{sec:Discussion}$).

\subsection{Model Description}

Following \citet{Phinney&Hansen93}, we model the disk evolution by a patchwork of matched self-similar solutions corresponding to each stage described in $\S\ref{sec:stages}$. The viscous time at the outer disk radius $R_{\rm d}(t)$ controls the rate of mass accretion rate $\dot{M}_{\rm d}(t)$ reaching smaller scales and hence the temporal spreading rate set by angular momentum conservation, as described in Appendix~\ref{sec:Appendix_SelfSimilar}.  Transitions between different regimes occur once either (1) the value of $R_{\rm d}$ exceeds the critical radii $R_{\rm rad}$, $R_{\rm irr}$, or $R_{\rm evap}$; (2) the disk becomes optically thin ($\tau = \sqrt{2/3}$); (3) the opacity regime at $R_{\rm d}$ changes (if the disk is in the radiative regime), or (3) the accretion rate becomes sub-Eddington (if the disk is in the irradiated phase).

We assume that the disk achieves a new self-similar solution instantaneously following any regime transition. As such, the complete solution is continuous in all variables, with two exceptions: (a) the transition from the unstable radiation-dominated regime to the gas pressure-dominated solution at the radiative disk transition. (b) the transition to a low temperature (neutral) solution at onset of the recombination instability.  In both cases, the disk discontinuously transitions to a lower $\dot{M}_{\rm d}$ regime in which the viscous timescale is larger than just prior to the transition.  The disk therefore stalls at these transitions for a time $t \sim t_{\rm visc}$, until the new self-similar evolution proceeds \citep[see also][]{Shen&Matzner14}.

The similarity solutions described in Appendix~\ref{sec:Appendix_SelfSimilar} provide the radial scaling of disk variables, which then allows us to extrapolate the solution at any given time from $R_{\rm d}$ to smaller radii. Using the same criteria discussed above, we determine the critical radii at which the disk transitions between different regimes, and extrapolate from any such transition radii inwards using the radial similarity solutions for the new state.

\section{Results} \label{sec:DiskResults}

Starting from the final configuration of WD-NS merger disks calculated by \citetalias{Margalit&Metzger16}, we follow the remainder of the disk evolution.  Our main goal is to investigate the disk conditions at $R_{\rm p} \sim 0.3-0.5$ AU where the PSR B1257+12 planets reside, and close to where they likely formed originally (\citealt{Hansen+09}).  We focus on the mass available for planet formation near this radius as compared to the $\simeq 8M_{\oplus}$ required to explain the pulsar planets, along with the local disk temperature.  Solid condensation and subsequent planet formation ($\S\ref{sec:PlanetFormation}$) begins only once the temperature has decreased to the condensation of gaseous carbon to graphite grains at a critical temperature of $T_{\rm c} \lesssim 2000 ~{\rm K}$ (\citealt{Goeres96}).  

We begin by describing a fiducial model, which overviews the basic stages of the disk evolution.  We then move on to a parameter survey that more thoroughly examines the requisite conditions for planet formation.

\subsection{Fiducial Model}
\label{sec:fiducial}

\begin{figure} 
\centering
\epsfig{file=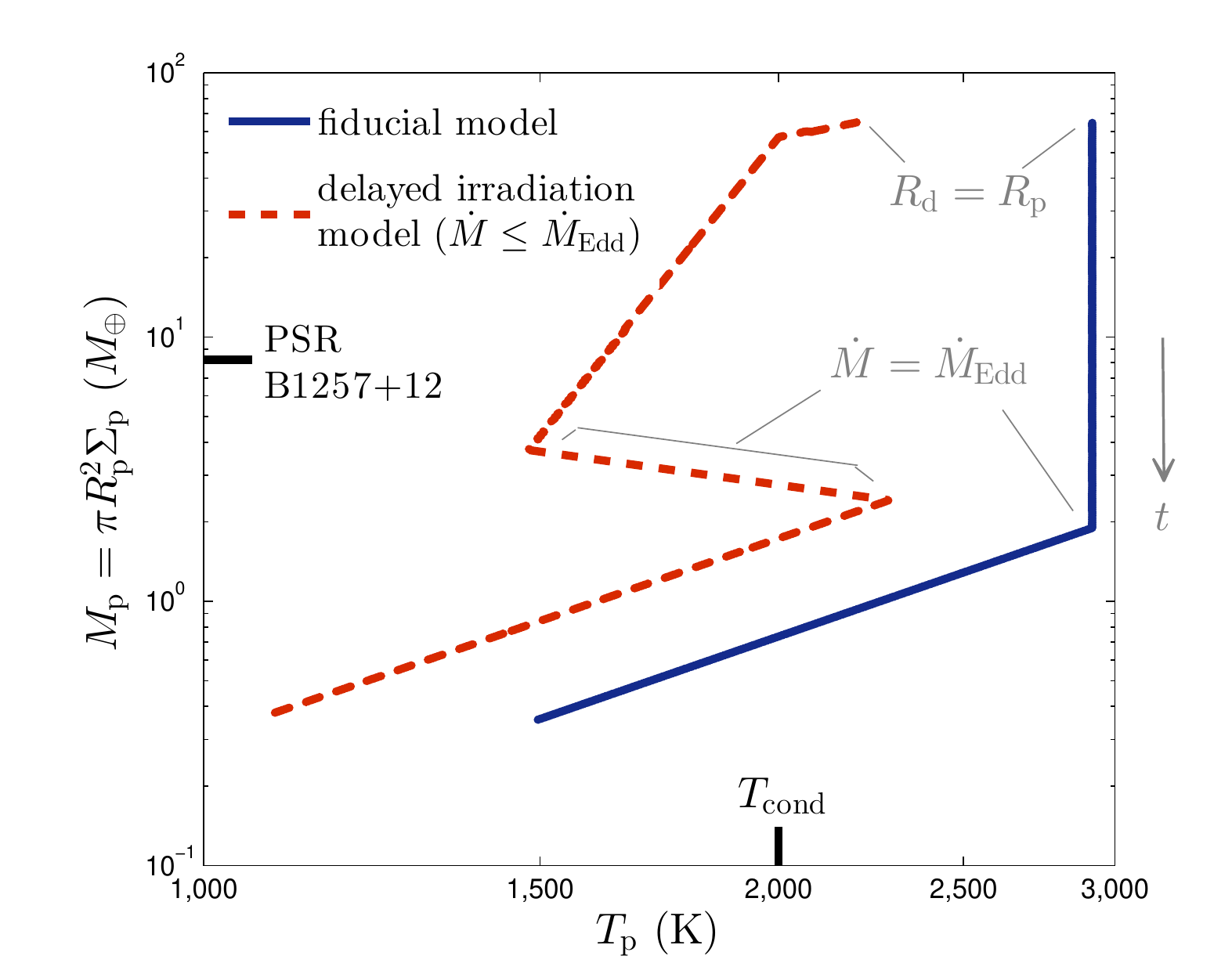,angle=0,width=0.5\textwidth}
\caption{Local mass and temperature at the presumptive planet forming radius, $R_{\rm p} = 0.4$ AU, for the fiducial model (solid) and a variation where irradiation of the outer disk only begins once the accretion rate becomes sub-Eddington (dashed line). The grey arrow indicates the temporal evolution direction. The disk first reaches $R_{\rm p}$ in the super-Eddington irradiated regime.  For the fiducial model, the temperature remains constant during this phase (equation~\ref{eq:Tp_irr_MEdd}), until the accretion rate drops below $\dot{M}_{\rm Edd}$ and the disk transitions to the sub-Eddington irradiated regime.  In the delayed irradiation model, the disk is viscously heated when it first reaches $R_{\rm p}$. Incident radiation heats the disk once $\dot{M}=\dot{M}_{\rm Edd}$ and briefly maintains the disk at a constant accretion rate, after which $\dot{M}$ decreases below $\dot{M}_{\rm Edd}$ and the solution evolves similarly to the fiducial model. The thick markings on the horizontal (vertical) axis mark the graphite condensation temperature (PSR B1257+12 combined planetary mass), respectively.} \label{fig:Mp_Tp}
\end{figure}

Our baseline model corresponds to the disruption of a $0.6 M_\odot$ C/O WD by a NS of mass $1.4 M_\odot$.  We assume that the WD composition is half carbon and half oxygen.
For simplicity, we take a constant mean molecular weight of $\mu \simeq 13$ (corresponding to the neutral atomic C/O phase) while calculating the disk evolution depicted in Figs.~\ref{fig:Evolution}-\ref{fig:Mp_Tp}. In practice, the mean molecular weight should be smaller, $\mu \simeq 1$, at earlier times when the gas is fully ionized, and larger, $\mu \simeq 28$, at late times once a significant fraction of the gas condenses into carbon-monoxide molecules.
We adopt a Shakura-Sunyaev viscosity parameter of $\alpha = 0.1$, and a wind mass-loss exponent of $p \simeq 0.44$ during the RIAF phase\footnote{In the \citetalias{Margalit&Metzger16} model, the value of $p$ is determined by the asymptotic velocity of the winds relative to the local escape speed of the disk and the value of the Bernoulli integral to which the midplane is regulated by wind cooling.}. 
The numerical calculations of \citetalias{Margalit&Metzger16} corresponding to these parameters conclude at $t \simeq 136~{\rm s}$, at which point $R_{\rm d}\simeq 5.8 \times 10^9~{\rm cm}$, $\Sigma_{\rm d} \simeq 1.5 \times 10^{12}~{\rm g~cm}^{-2}$, and $\theta_{\rm d} \simeq 0.43$. A total of $0.34 M_\odot$ has been lost from the disk by this early time, due to disk winds and accretion onto the NS.  Finally, in our fiducial solution we assume that the outer disk is irradiated by the inner accretion flow, even during the super-Eddington phase.  

Fig.~\ref{fig:Evolution} shows the temporal evolution of the outer disk radius (top panel), surface density (middle), and temperature (bottom panel), for this fiducial model. Vertical dashed curves indicate key transitions in the outer disk's accretion regime, and vertical dotted curves demarcate transitions in the opacity law at $R_{\rm d}$. 
The disk initially evolves in the RIAF phase, until $t \simeq 4 \times 10^5~{\rm s}$ when the disk transitions to a radiatively cooled regime ($R_{\rm d} = R_{\rm rad}$). This transition is discontinuous in temperature and scale height due to onset of the Lightman-Eardley instability, following which the disk collapses to the geometrically-thin, gas pressure-dominated solution.  The disk stalls at this transition radius until $t$ equals the new (larger) viscous timescale, after which the self-similar radiative evolution commences. 

At $t \simeq 5 \times 10^8~{\rm s}$ the temperature decreases to $8000~{\rm K}$ and the disk encounters the recombination instability, transitioning discontinuously to the lower temperature solution in the opacity gap. Once again, a brief stalling in the evolution is associated with this sudden transition. Shortly thereafter, $R_{\rm d} = R_{\rm irr}$ and the outer disk becomes irradiation heated. Irradiation is dominant throughout the remainder of the disk evolution, and is only affected by the transition from a super-Eddington to sub-Eddington accretion rate around $t \sim 2 \times 10^{10}~{\rm s}$. The evolution finally terminates once $R_{\rm d} = R_{\rm evap}$, following which the disk photo-evaporates on a viscous timescale.

The conditions of the gaseous disk near the radius of presumptive planet formation for this baseline model are illustrated in Fig.~\ref{fig:Mp_Tp}, which shows the local mass and midplane temperature at $R_{\rm p} = 0.4$ AU.  Initially, $R_{\rm d} < R_{\rm p}$ and thus no appreciable mass is present at this location.  When the outer disk radius first crosses $R_{\rm p}$, the mass at this position is $\simeq 65 M_\oplus$, but the temperature is marginally higher than required for solid condensation, $T = T_{\rm c} \simeq 2000~{\rm K}$. As time elapses (grey arrow), the disk continues to spread, accreting some of its mass, and carrying the remainder to larger radii, so the local mass at $R_{\rm p}$ decreases.
The temperature at this fixed radius remains constant for some period, as the disk is locally in the super-Eddington irradiation heated regime (equation~\ref{eq:Tp_irr_MEdd}). Only later, once the accretion rate becomes sub-Eddington, does the local temperature decrease to the point to allow solid condensation.

Evidently, by the time the temperature at $R_{\rm p}$ decreases to $2000~{\rm K}$, the local disk mass available for planet formation is an order of magnitude too low.  At face value, this rules out our fiducial model as the source of planets in the PSR B1257+12 system.  However, if we instead consider the variation of the fiducial model described earlier, in which the irradiation phase is delayed until the accretion rate becomes sub-Eddington (dashed line in Fig.~\ref{fig:Mp_Tp}), then planet formation can commence at an earlier phase, when the disk mass is still $\sim 60M_{\oplus}$.  Although $T_p$ again temporarily increases above the condensation temperature once irradiation turns on at $\dot{M} = \dot{M}_{\rm Edd}$, this may not destroy the solids that have formed because the sublimation and melting temperature of solid carbon is higher than the condensation temperature.  

We now discuss the broader range of disk parameters that allow the mass-temperature constraints needed for planet formation to be satisfied.

\subsection{Analytic Considerations}

The two main free parameters in our model are the strength of the viscosity $\alpha$ and the wind mass-loss exponent $p$.  Some uncertainty also arises due to our approximate opacity curve, but this has little affect on the results if the disk is in the irradiation phase at the time of solid condensation.  The initial conditions are also reasonably well determined:  the disk aspect ratio has a fixed value $\theta \sim 0.4$ set by its  initial thermal energy determined by energy conservation during the merger.  The initial size and surface density of the disk are solely determined by WD mass (\citetalias{Margalit&Metzger16}), which varies by less than a factor of $2$ \citep[e.g.][]{Liebert+05}.  

In order to explore a wider parameter space than the specific models studied by \citetalias{Margalit&Metzger16}, we consider the initial disk conditions at yet earlier times of minutes-hours, near the time when the accretion rate first peaks following the WD disruption and before wind mass loss has become significant.  Specifically, we consider $R_{\rm d, 0} = 2 \times 10^9~{\rm cm}$, $\Sigma_{\rm d, 0} = 2 \times 10^{13}~{\rm g~cm}^{-2}$, $\theta_{\rm d, 0} = 0.4$, for an initial disk mass of $M_{\rm d, 0} = 0.6 M_\odot$. These values, informed by our previous numerical studies, describe the initial disk well.

\subsubsection{Analytic Estimates of Mass and Temperature at Planet Formation Radius}

As illustrated by our fiducial model, the available mass budget at the planet formation radius $R_{\rm p}$ peaks when the outer disk radius first reaches this location (Fig.~\ref{fig:Mp_Tp}).  We thus focus on estimating the disk mass at this time using global angular momentum considerations.  In doing so we must treat the RIAF and the radiatively cooled regimes separately.  

The RIAF phase ends when the disk first becomes radiative ($R_{\rm d} = R_{\rm rad}$). 
Assuming electron scattering dominates the opacity during this early phase, then $R_{\rm rad} \propto \tau^2 \propto \Sigma_{\rm d}^2$ (eq.~\ref{eq:R_rad}).  During the RIAF phase, the disk radius evolves according to equation (\ref{eq:RdMd_RIAF}).  At the radiative transition, the disk mass has therefore decreased to a fraction 
$\left( {R_{\rm rad, 0}}/{R_{\rm d, 0}} \right)^{-(2p+1)/(4p+12)}$ 
of its initial value, where \begin{align} \label{eq:R_rad0}
R_{\rm rad, 0} &\equiv \left(\frac{3}{2}\right)^4 \eta_{\rm rad}^2 \alpha^2 \theta_{\rm d,0}^2 \bar{\kappa}_{\rm es}^2 \Sigma_{\rm d, 0}^2 \frac{G M_{\rm NS}}{c^2} \\ \nonumber
&\approx 6.7 \times 10^{25}\,{\rm cm} \left(\frac{\alpha}{0.01}\right)^2 \left(\frac{\Sigma_{\rm d,0}}{2\times10^{13}\,{\rm g\,cm}^{-2}}\right)^2
~.
\end{align}
is the initial value of the radiative transition radius,
and in the second equality we assume canonical values of $\eta_{\rm rad} = 0.5$, $\theta_{\rm d,0}=0.4$ and $M_{\rm NS} = 1.4 M_\odot$.

After the radiative transition, the disk evolves without further angular momentum loss, expanding according to equation~(\ref{eq:RdMd_radiative}), and losing additional mass solely to accretion.  The remaining mass once the disk first reaches $R_{\rm p}$ is thus
\begin{equation} \label{eq:M_dp}
M_{\rm d}\left( R_{\rm d} = R_{\rm p} \right) = M_{\rm d, 0} \left(\frac{R_{\rm p}}{R_{\rm d, 0}}\right)^{-1/2} \left(\frac{R_{\rm rad, 0}}{R_{\rm d, 0}}\right)^{-p/(2p+6)}.
\end{equation}
This mass is notably significantly smaller than the standard result for a spreading disk under the assumption of zero wind mass loss ($p = 0$; e.g. \citealt{Phinney&Hansen93, Currie&Hansen07}).  Finally, the total disk mass, $M_{\rm d}$, is related to the local mass at the outer radius via
\begin{equation} \label{eq:A}
A \pi R_{\rm d}^2 \Sigma_{\rm d} = M_{\rm d} ,
\end{equation}
where $A \approx 3$ is a constant determined by an analysis of our numerical results \citep[see Appendix~A of][]{Metzger+08}.

The temperature at $R_{\rm p}$ depends on local energy balance. Three separate cases require consideration based on the evolutionary phases described in \S\ref{subsec:EvolutionStages}.

For an irradiated disk during the super-Eddington phase, the temperature at the planet formation radius is constant
\begin{equation} \label{eq:Tp_irr_MEdd}
T_{\rm p}^{\rm (irr, Edd)} \simeq 2600 \,{\rm K} \left(\frac{R_{\rm p}}{0.4\,{\rm AU}}\right)^{-3/7} \left(\frac{\mu}{28}\right)^{-1/7},
\end{equation}
where we have used equations~(\ref{eq:qdot_irr}) and (\ref{eq:qdot_rad}) with $f(\tau)=1$.  Here we adopt $\mu = 28$ appropriate if most of the disk mass is locked up in molecular CO, as expected at temperatures $\lesssim 4000\,{\rm K}$ (see \S\,\ref{sec:PlanetFormation}).

Similarly, for an irradiated disk during the sub-Eddington phase, the temperature at $R_{\rm p}$ is given by
\begin{align} \label{eq:Tp_irr}
&T_{\rm p}^{\rm (irr)} \simeq \left[ \frac{\alpha M_{\rm p}}{R_{\rm p}^{2}} \frac{27}{392 \pi \sigma R_{\rm NS}} \left(\frac{k_{\rm B}}{\mu m_{\rm p}}\right)^{3/2} \right]^{2/5}  \\ \nonumber
&\approx 1000\,{\rm K}\left(\frac{\alpha}{0.01}\right)^{2/5}\left(\frac{M_{\rm p}}{10M_{\oplus}}\right)^{2/5} \left(\frac{R_{\rm p}}{0.4\,{\rm AU}}\right)^{-4/5} \left(\frac{\mu}{28}\right)^{-3/5},
\end{align}
where $M_{\rm p} \equiv \pi R_{\rm p}^2 \Sigma(R_{\rm p})$ is the {\it local} mass at radius $R_{\rm p}$.  In deriving the accretion luminosity used to calculate the irradiation heating from equation (\ref{eq:qdot_irr}), we have expressed the local accretion rate as 
$\dot{M} = 2\pi r \Sigma |v_r| = 9 \alpha \theta^2 \Omega M_{\rm p} /7$,
where the factor $9/7$ is specified by the self-similar solution in this regime.

Finally, we consider a third scenario in which the disk first reaches $R_{\rm p}$ during the viscously heated phase.  Although the temperature in this phase is not easily tractable in general due to the complicated form of $\bar{\kappa}(\rho, T)$, we focus on a simple and common case in which the disk is in the opacity-gap region of the opacity curve (Appendix~\ref{sec:Appendix_Opacity}).  In this temperature range of $2000\,{\rm K} \lesssim T \lesssim 8000\,{\rm K}$, where the gas is mainly neutral but not yet cool enough to form dust, we approximate the opacity by a constant low value of $\bar{\kappa}_{\rm gap} = 10^{-2} ~{\rm cm}^2~{\rm g}^{-1}$.

Using equations (\ref{eq:qdot_visc}) and (\ref{eq:qdot_rad}) along with equation~(\ref{eq:f_of_tau}) in the optically thick regime, we find that the disk temperature at $R_{\rm p}$ in the viscously heated regime, is given by
\begin{align} \label{eq:Tp_visc}
T_{\rm p}^{\rm (visc)} &\simeq \left[ \frac{\alpha M_{\rm p}^2}{ R_{\rm p}^{11/2}} \frac{27 \bar{\kappa}_{\rm gap}}{32 \pi^2 \sigma} \frac{k_{\rm B}}{\mu m_{\rm p}} \sqrt{G M_{\rm NS}} \right]^{1/3} \\ \nonumber
&\approx 105\,{\rm K} \left(\frac{\bar{\kappa}_{\rm gap}}{10^{-2}\,{\rm cm^{2}\,g^{-1}}}\right)^{1/3}\left(\frac{\alpha}{0.01}\right)^{2/5} \left(\frac{M_{\rm p}}{10M_{\oplus}}\right)^{2/3} \\ \nonumber
&~~~~~~~\times \left(\frac{R_{\rm p}}{0.4\,{\rm AU}}\right)^{-11/6} \left(\frac{\mu}{28}\right)^{-1/3}.
\end{align}
Note that this expression is only applicable for temperatures above the dust-condensation temperature $T_{\rm c} \sim 2000\,{\rm K}$.  Grain opacity becomes dominant once dust formation commences, increasing the value of $\bar{\kappa}$. The resulting strong dependence of $\bar{\kappa}$ on temperature regulates the latter to a fixed value $\sim T_{\rm c}$ across a wide range of densities.

\subsubsection{Parameter Study}

\begin{figure} 
\centering
\epsfig{file=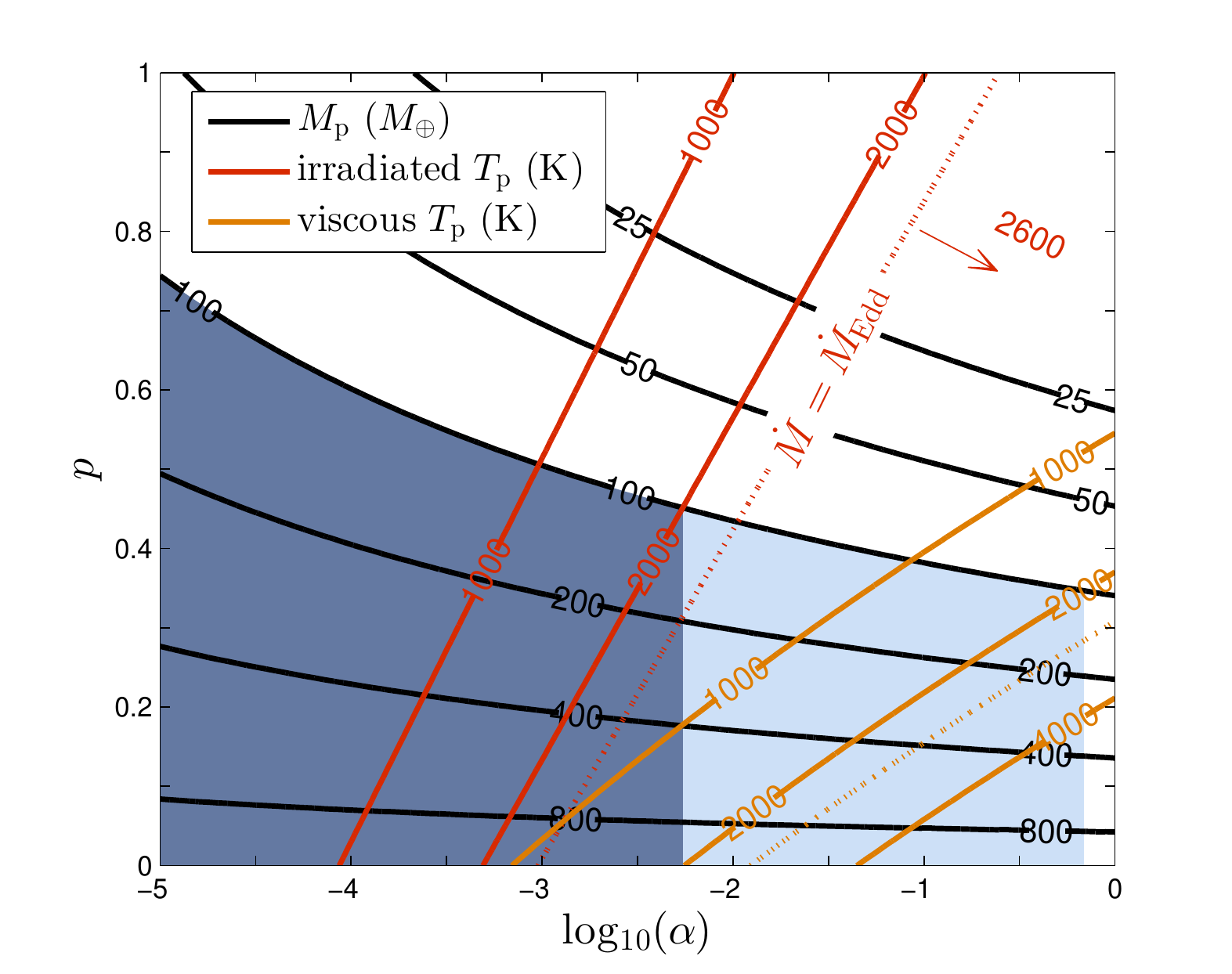,angle=0,width=0.5\textwidth}
\caption{Mass and temperature conditions when the disk first spreads to the planet formation radius $R_{\rm d}=R_{\rm p} = 0.4\,{\rm AU}$ in the parameter space of the viscosity ($\alpha$) and wind mass-loss exponent ($p$). Black contours show the local mass at $R_{\rm p}$, $M_{\rm p}$ (labeled in units of $M_\oplus$) calculated from equation~(\ref{eq:M_dp}) assuming $A=3$ (see equation~\ref{eq:A}). The local temperature at the same position, $T_{\rm p}$ is labeled in units of Kelvin and plotted in red (brown) for an irradiation (viscously)-heated disk, as calculated via equations~(\ref{eq:Tp_irr_MEdd}-\ref{eq:Tp_visc}). Larger values of $p$ result in strong initial outflows which therefore decrease the available mass at late times, whereas lower values of $\alpha$ decrease the viscous heating and accretion rate, leading to smaller temperatures. The shaded blue regions show the allowed parameter space if a mass of $100 M_{\oplus}$ when the temperature first reaches $2000\,{\rm K}$ is required to produce the observed pulsar-planets assuming a formation efficiency of $8\%$ (\S \ref{sec:PlanetFormation}). 
Despite this conservative assumption, a reasonably wide parameter-space satisfies these constraints.} \label{fig:ParameterSpace}  
\end{figure}

We now use our derived expressions for the disk mass $M_{\rm d}$ and temperature once the disk first reaches the planet formation radius $R_{\rm p} = 0.4$ AU to constrain the parameter space required for planet formation.  Fig.~\ref{fig:ParameterSpace} shows with black contours the local disk mass at the planet formation radius, $M_{\rm p}/M_\oplus$, in the space of viscosity $\alpha$ and RIAF wind mass loss parameter $p$.  Also shown with red (brown) curves are contours of constant temperature in K for the irradiation (viscously)-heated regimes.  The viscous regime temperature curves are calculated assuming a constant opacity of $\bar{\kappa} = 10^{-2}\,{\rm cm}^2\,{\rm g}^{-1}$ (equation~\ref{eq:Tp_visc}).

For large values of $p$ and small values of $\alpha$ (upper left corner of Fig.~\ref{fig:ParameterSpace}), when the disk first reaches $R_p$ it is irradiation-heated and the accretion rate is sub-Eddington, such that equation (\ref{eq:Tp_irr}) applies.  Below the dotted red curve ($\dot{M}=\dot{M}_{\rm Edd}$) the accretion rate is super-Eddington at $R_{\rm p}$.  For small values of $p$ and large values of $\alpha$ (bottom right corner of Fig.~\ref{fig:ParameterSpace}), the disk reaches $R_{\rm p}$ in the viscously-heated radiative phase (equation~\ref{eq:Tp_visc}).  The transition between the viscously-heated and irradiation-heated regimes for the fiducial model is shown with a dotted brown curve, along which $T_{\rm p}^{\rm (visc)} = T_{\rm p}^{\rm (irr, Edd)}$ (equations~\ref{eq:Tp_irr_MEdd},\ref{eq:Tp_visc}). Thus, between the dotted curves, the disk is irradiation-heated and accreting above the Eddington limit and the temperature is constant, $T = T_{\rm p}^{\rm (irr, Edd)} \simeq 2600\,{\rm K}$ (equation~\ref{eq:Tp_irr_MEdd}). In the delayed-irradiation model, the disk instead remains viscously heated until $\dot{M} \lesssim \dot{M}_{\rm Edd}$, such that the temperature is less than $2600\,{\rm K}$ in between the two dotted lines.

In reality, the disk temperature in the viscous-heated regime below the condensation temperature $T_{\rm c} \sim 2000\,{\rm K}$ will differ from those depicted in Fig.~\ref{fig:ParameterSpace} due to the increase in opacity that accompanies grain formation.  Once a supply of small carbonaceous grains becomes available, rapid growth by coagulation onto larger grains should proceed efficiently, rapidly forming larger solids which sink to the disk midplane, while simultaneously depleting and regulating the number of small grains that contribute most to the opacity \citep[e.g.][]{Dullemond&Dominik05}.  Although detailed modeling of this process is beyond the scope of this paper, the net effect of grain formation is to regulate the disk temperature to a value near the condensation temperature for a wide range of densities (i.e. independent of $M_{\rm p}$).  This will extend the parameter space with $T \lesssim 2000\,{\rm K}$ to lower values of $\alpha$ and higher values of $p$ than predicted by our constant opacity model. 

The shaded blue regions in Fig.~\ref{fig:ParameterSpace} show the allowed parameter-space to form the PSR B1257+12 planets, under the assumption that a local disk mass of $100 M_\oplus$ is required at the time when the temperature first drops below $2000\,{\rm K}$ (corresponding to a planet formation efficiency of 8$\%$; see $\S\ref{sec:PlanetFormation}$). The dark blue area constrains the fiducial model, whereas the underlying lightly shaded region applies to the irradiation-delayed model. Both are limited from above by the $M_{\rm p} = 100 M_\oplus$ contour, and terminate at the maximal value of $\alpha$ where this mass contour meets the $T_{\rm p} = 2000\,{\rm K}$ curve.
The allowed parameter-space includes cases in which the disk first reaches $R_{\rm p}$ at temperatures $\gtrsim 2000\,{\rm K}$. These models are still viable if enough mass remains near $R_{\rm p}$ once the disk further expands and cools below the condensation threshold (as heuristically pictured in Fig.~\ref{fig:Mp_Tp}).  Since equations~(\ref{eq:Tp_irr_MEdd}-\ref{eq:Tp_visc}) do not depend explicitly on $p$, the boundaries of the permitted parameter-space are vertical lines.

Favorable conditions for planet formation within the fiducial model require low values of $\alpha \lesssim 6 \times 10^{-3}$, due to the lower temperature of the disk when the accretion rate is sub-Eddington.  The simplest version of the delayed-irradiation model permits larger values of $\alpha$ since in this case the disk first reaches $R_{\rm p}$ in the cooler, viscously-heated phase.  The parameter-space constraints are further alleviated if the planet-formation efficiency is larger than the value of 8$\%$ we have assumed. Even relaxing the efficiency moderately to $\sim 16\%$ (see \S \ref{sec:PlanetFormation}), would expand the allowed parameter-space up to the $M_{\rm p} = 50 M_\oplus$ contour.  This would permit viscosities up to $\alpha \approx 10^{-2}$ for the fiducial model and values of $p \sim 0.5$ favored by numerical simulations of RIAFs for both fiducial and delay-irradiation models.

A time-varying value of $\alpha$, as might be expected as the disk ionization state changes, can also increase the allowed parameter-space.  The mass contours in Fig.~\ref{fig:ParameterSpace} depend on $\alpha$ only through the value of $R_{\rm rad,0}$ (equation~\ref{eq:R_rad0}), which determines the radiative transition time at early times when the disk is hot and fully ionized. By contrast, the temperature contours of Fig.~\ref{fig:ParameterSpace} are determined by the value of $\alpha$ at late times, after the disk material has largely recombined.  If the lower ionization state of the disk reduces the effective value of $\alpha$, e.g. due to suppression of the MRI by non-ideal MHD effects, then it would become easier to simultaneously satisfy both mass and temperature constraints on planet formation. However, we note that it is unlikely that the disk will become entirely `dead', since even the low ionization levels of trace alkaline elements are sufficient to sustain the MRI \citep{Gammie96, Armitage10}.  X-ray irradiation from the inner accretion flow will also maintain a significant ionized column on the disk surface.

\subsubsection{Application to Supernova Fall-back Disks}
Although our focus is on the WD-NS merger scenario, we can apply our estimates to show why the supernova fallback model is disfavored for pulsar planet formation.  The expected angular momentum and mass of such disks are at most\footnote{In fact, \cite{Perna+14} find that disk formation is disfavored altogether for single star stellar evolution models which include commonly used prescriptions for interior angular momentum transport due to magnetic torques resulting from the Spruit-Taylor dynamo.} $J_{\rm d,0} \sim 10^{49}\,{\rm erg\,s}$ and $M_{\rm d,0} \approx 10^{-3}-10^{-1} M_\odot$ \citep{Chevalier89}.  The factor of $\lesssim 10^{-2}$ smaller angular momentum in the fallback scenario compared with the WD-NS merger case severely limits the remaining mass reservoir at the planet forming radius.  

Even neglecting disk winds, the local mass reaching $R_{\rm p}$ (for $A=3$) is only
\begin{equation} \label{eq:Mp_SN_upperbound}
M_{\rm p} < 17 M_\oplus \left(\frac{J_{\rm d,0}}{10^{49}\,{\rm erg\,s}}\right) \left(\frac{R_{\rm p}}{0.4\,{\rm AU}}\right)^{-1/2} ,
\end{equation}
in tension with the observed PSR B1257+12 planetary system unless the metallicity of the accreting gas is very high. 

In reality, unbound winds during the early RIAF phase are challenging to avoid (e.g.~\citealt{MacFadyen&Woosley99}).  The actual mass reaching $R_{\rm p}$ is thus smaller than the above upper-limit by a factor of $(R_{\rm rad, 0} / R_{\rm d, 0})^{-p/(2p+1)}$ (eq.~\ref{eq:M_dp}).
For canonical parameters and a wind mass-loss exponent of $p=0.5$, this further reduces the gaseous mass by an order of magnitude,
\begin{align} \label{eq:Mp_SN}
M_{\rm p} &\underset{p = 0.5}\approx 1.6 M_\oplus \left(\frac{J_{\rm d,0}}{10^{49}\,{\rm erg\,s}}\right)^{12/7} \left(\frac{M_{\rm d,0}}{10^{-2} M_\odot}\right)^{-6/7} \\ \nonumber
&\times \left(\frac{\alpha}{0.01}\right)^{-1/7} \left(\frac{R_{\rm p}}{0.4\,{\rm AU}}\right)^{-1/2} \leq {\rm eq.~(\ref{eq:Mp_SN_upperbound})}, 
\end{align}
inconsistent with even an $100\%$ planet-formation efficiency model (note that for any range of parameters, equation~\ref{eq:Mp_SN} must be truncated from above by equation~\ref{eq:Mp_SN_upperbound}).
We conclude that low-angular momentum disks, such as those anticipated in most core collapse events, cannot channel enough mass to radii $\approx 0.4\,{\rm AU}$ to explain the PSR B1257+12 planetary system.

The debris disk detected from its infrared excess around the young millisecond pulsar 4U 0142+61 provides possible evidence for the existence of fallback disks \citep{Wang+06}.  However, the inferred disk radius $R_{\rm d} \simeq 0.04\,{\rm AU}$ and mass $M_{\rm d} \sim 10\,M_\oplus$ in this case imply a modest angular momentum of $J_{\rm d} \sim 5 \times 10^{47}\,{\rm erg\,s}$, consistent with our estimates for the effects of mass loss during an early RIAF phase as given by equation (\ref{eq:Mp_SN}).

\section{Planet Formation Scenario} \label{sec:PlanetFormation}

For realistic parameters, we find that the mass in the disk when it reaches $R_{\rm p}$ greatly exceeds the combined masses of the observed PSR B1257+12 planets $\simeq 8 M_\oplus$.  This is especially true if the outer disk is shielded from irradiation by the inner disk during the super-Eddington phase.  In the following, we outline a possible formation mechanism for the pulsar planets which differs in several respects from the `standard' planetary formation scenario. An entirely separate and self-contained work could be devoted to this complex issue, yet here we only propose a general scenario for this process and point-out some of the key issues involved.

A unique aspect of planet formation in our scenario is the unusual C/O-dominated composition.  The innermost regions of the disk undergo nuclear burning at early times in the RIAF phase, synthesizing intermediate mass and iron group elements (\citealt{Metzger12}; \citetalias{Margalit&Metzger16}).  However, these inner layers are either unbound from the system by winds or accreted by the NS, such that the composition of the outermost disk remains dominated by unburned C/O.\footnote{However, the possibility cannot be excluded that moderate amounts of heavier elements synthesized at small radii reach the upstream, either due to radial turbulent diffusion within the disk or due to the fall-back of wind ejecta launched from the inner disk.}  

The $Z=1$ metallicity of our proto-planetary disk would suggest a very efficient planet-formation process, since an order unity of the available disk mass could condense into solids.
Upon closer examination, while carbonaceous dust grains can form from gaseous carbon, predominantly as graphite (there is no hydrogen from which PAHs could form), most of the oxygen is effectively inert because it must combine with other intermediate mass elements to form silicate grains.  

Furthermore, a substantial fraction of the carbon mass reservoir will inevitably be trapped in carbon-monoxide (CO), which is stable and forms at a higher temperature
than graphite condensation.  In particular, using the thermal CO formation/destruction rates of \cite{Lazzati&Heger16}, we find that carbon-monoxide condenses at $T_{\rm CO} \sim 4000\,{\rm K}$ for characteristic densities of $n_{14} = n / 10^{14}\,{\rm cm}^{-3}$, and which in detail is well fit by the formula $T_{\rm CO}/{\rm K} \approx 4124 + 304 \log_{10}(n_{14}) + 21 \log_{10}(n_{14})^2$.

The formation of CO at temperatures higher than solid condensation suggests that the solid formation efficiency could be much lower than the maximum $\sim 100\%$ efficiency allowed for unity metallicity.  Two ingredients control the amount of carbon that will be trapped in CO versus that available to condense into solids. The first is the initial C/O number ratio of our disk, which is essentially that of the initially disrupted WD.   The WD C/O ratio is generally believed to be close to unity, with increasing oxygen abundances with larger WD mass.
Still, current stellar evolution model estimates of WD composition are highly uncertain due to underlying uncertainties in the nuclear physics, primarily in the $^{12}$C$(\alpha,\gamma)^{16}$O reaction rate. \cite{Fields+16} recently applied a Monte Carlo approach to estimate the uncertainties in WD central carbon and oxygen abundances, finding that $\Delta X_{^{12}{\rm C}} \approx \Delta X_{^{16}{\rm O}} \approx 0.4$, indicating that the C/O ratio is essentially unconstrained.
Given these uncertainties, it is plausible that $X_{^{12}{\rm C}} \gtrsim X_{^{16}{\rm O}}$ yielding C/O ratios of order $1.1$ or larger. This would suggest a $\sim 10\%$ or greater efficiency in the condensation of solids.

Even if the C/O ratio is below unity, carbonaceous dust formation may not be inhibited.  UV/X-ray irradiation from the inner disk will photodissociate CO molecules in the upper disk layers, continuously generating a fresh supply of free carbon. The fate of these carbon monomers depends sensitively on the ambient conditions. If the rate of CO formation exceeds the rate of carbonaceous dust condensation, then recently freed carbon will immediately capture onto nearby oxygen before sinking to the midplane.  However, the rate of thermal CO formation decreases exponentially with the declining disk temperature (\citealt{Lazzati&Heger16}), such that as the disk continues to cool solid formation may eventually come to eclipse CO formation.  This could occur even in regions of the disk which remain shielded by X-rays and hence are conducive to the required gas phase chemistry.  

Once carbon solids condense out of the gas phase, they will sink to the disk midplane and grow to large sizes by two body collisions.  The stage by which larger planetesimals grow beyond this point is uncertain.  One possibility is the direct collapse of the solid disk into gravitationally bound entities due to gravitational instability on a dynamical time (\citealt{Goldreich&Ward73}).  Vertical shear between the gaseous and solid disks induces turbulence, which can inhibit the gravitational instability (\citealt{Weidenschilling95}).  
However, \cite{Youdin&Shu02} have shown that sufficiently large ratios of the solid to gas surface densities inhibit such turbulence.
Specifically, the critical mass fraction of solids required for the gravitational instability to act effectively is (within factors of order unity) given by \citep{Sekiya98,Youdin&Shu02}
\begin{equation}
X_{\rm solids} \gtrsim \theta_{\rm gas} \approx 0.014 \left(\frac{T}{2000\,{\rm K}}\right)^{1/2} \left(\frac{R_{\rm p}}{0.4\,{\rm AU}}\right)^{1/2} .
\end{equation}
As discussed above, the C/O-dominated disks envisioned in our current scenario should easily satisfy this criterion.

The most unstable wavelength by which the gravitational instability proceeds is
\begin{equation}
\lambda \sim \left. \frac{2 \pi^2 G \Sigma }{ \Omega^2 } \right|_{\rm solids} \approx 3\times 10^{8}\,{\rm cm} \left(\frac{M_{\rm solids}}{10 M_\oplus}\right) \left(\frac{R_{\rm p}}{0.4\,{\rm AU}}\right) ,
\end{equation}
implying a characteristic (maximal) planetesimal mass of
\begin{equation} \label{eq:m_planetesimal}
m \sim 2\pi \Sigma R_{\rm p} \lambda 
\approx 3 \times 10^{-4} M_\oplus \left(\frac{M_{\rm solids}}{10 M_\oplus}\right)^2  .
\end{equation}
The remaining `rubble pile' must therefore undergo $\sim 10^4$ collisions to build up the earth mass planets observed in the PSR B1257+12 system. As pointed out by \cite{Miller&Hamilton01}, these collisions are unlikely to eject bodies from the system, since the escape velocity from such planetesimal embryos' surface, $\lesssim 1\,{\rm km\,s}^{-1}$ (equation~\ref{eq:m_planetesimal}), is significantly below the system escape speed $\sim 50\,{\rm km\,s}^{-1}$. 
This suggests that subsequent collisional assembly of the earth mass planets may be quite efficient, necessitating only $M_{\rm solids} \approx 10 M_\oplus$.

\section{Discussion} \label{sec:Discussion}

\begin{figure}
\centering
\epsfig{file=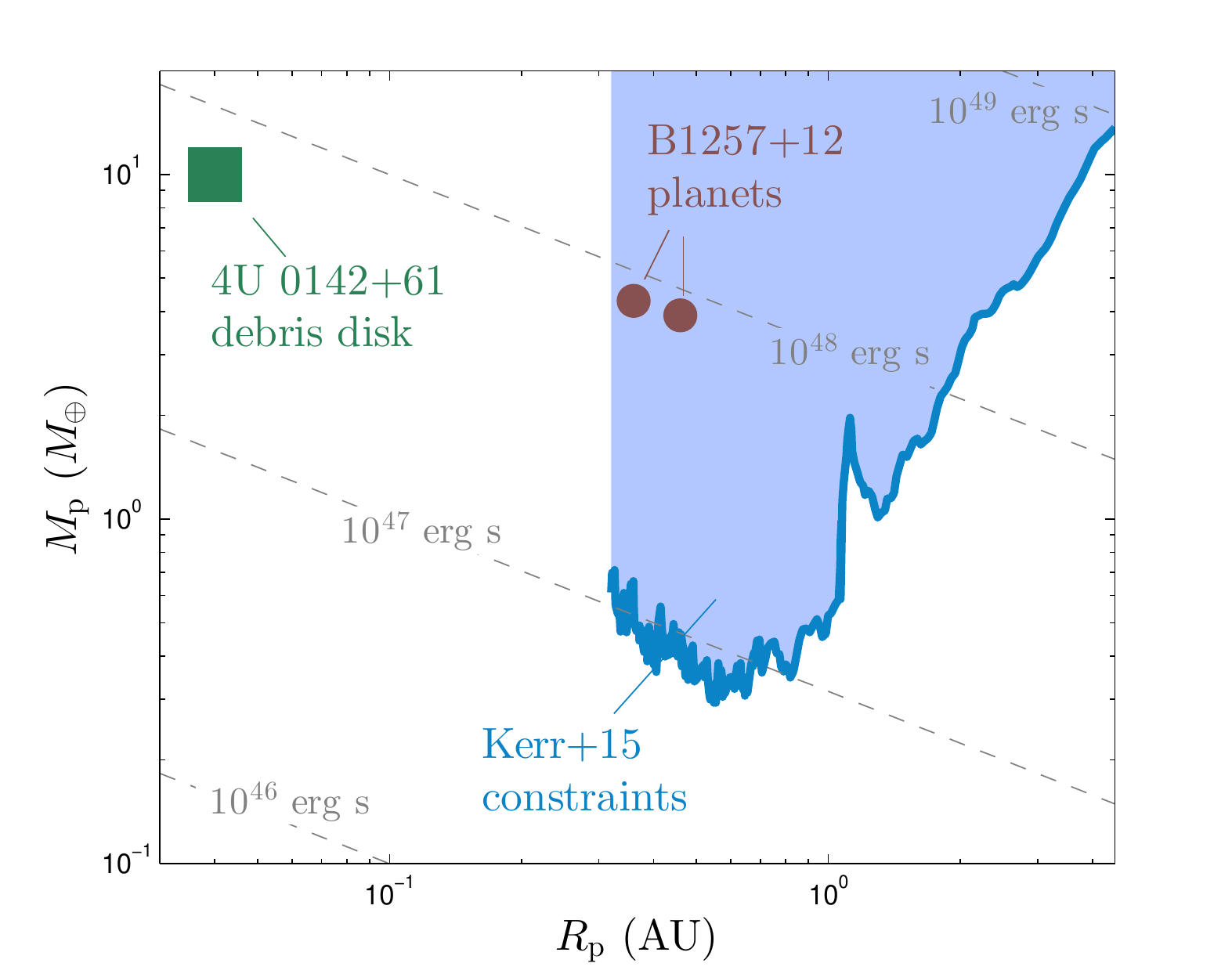,angle=0,width=0.5\textwidth}
\caption{Observational constraints on NS planetary systems. Pulsar planets of mass $M_{\rm p} \sin i$ at radii $R_{\rm p}$ are ruled out with $95\%$ confidence limits in the blue shaded region for a large sample of young pulsars \citep{Kerr+15}. Brown circles show the earth-mass PSR B1257+12 planets considered in this work \citep{Konacki&Wolszczan03}, while the green square shows a best-fit model for the 4U 0142+61 pulsar debris disk \citep{Wang+06}, a SN fall-back disk candidate. Lines of constant angular momentum are plotted as dashed grey curves.
The dearth of observed pulsar planetary systems, $\lesssim$1:150, is consistent with the expected low rate of WD-NS mergers, $\sim 10^{-2}$ of the core-collapse SN rate.} \label{fig:KerrConstraints}
\end{figure}

Compared to previous work modeling the time-dependent disk evolution (\citealt{Phinney&Hansen93,Currie&Hansen07}) we 
begin with initial conditions for the WD-NS merger scenario motivated by \citetalias{Margalit&Metzger16}, and
include for the first time the important effects of wind mass loss during the RIAF phase at early times.  
We identify a range of parameters which are consistent with necessary conditions for planet-formation around PSR B1257+12, and briefly discuss key aspects of the formation process unique to this scenario.
Finally, we show that including the early RIAF phase of mass loss in `low angular momentum' models, such as supernova fall-back disks, significantly reduces the gaseous disk mass, disfavoring these models for the PSR B1257+12 system. This would help explain the striking lack of planetary systems around the vast majority of pulsars, as summarized in Fig.~\ref{fig:KerrConstraints} adapted from \cite{Kerr+15}.  By contrast, the rate of observed pulsar planetary systems agrees well with the (albeit uncertain) rate estimates of WD-NS mergers (\citealt{OShaughnessy&Kim10,Kim+15,Bobrick+16}).

Though we have considered only C/O WDs, more massive O/Ne WDs can likewise merge with a binary NS companion, with rates that may even greatly exceed the C/O WD - NS merger rate \citep{Bobrick+16}. Our analysis can be applied to this scenario as well, with the exception that silicate dust (as apposed to graphite) will be the primary means of solid formation. Since the amount of silicates is limited by their $\sim$solar trace abundances, the planet-formation efficiency in this scenario drops and $\sim 10$ times more mass is required at $R_{\rm p}$. This severely constrains the prospects of an O/Ne WD merger as the progenitor of the PSR B1257+12 planetary system.

\begin{figure}
\centering
\epsfig{file=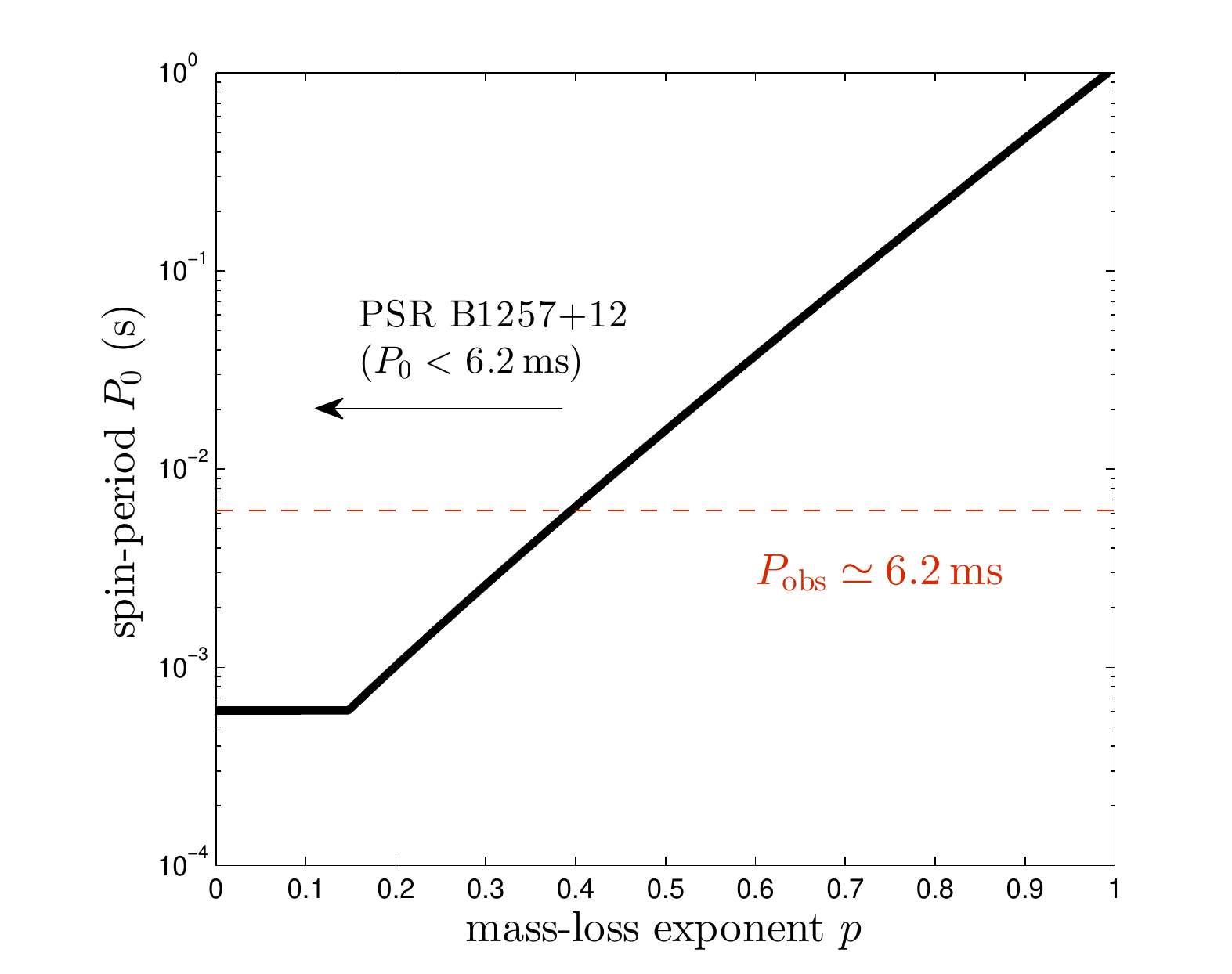,angle=0,width=0.5\textwidth}
\caption{Pulsar spin period following WD-NS merger accretion phase, $P_0$, versus the RIAF phase mass-loss exponent, $p$ (equation~\ref{eq:P_spin}). More mass reaches the NS surface for low values of $p$, leading to more significant spin-up, until the NS rotates at breakup frequency $\approx \Omega_{\rm k}(R_{\rm NS})$. The dashed red curve indicates the currenty observed PSR B1257+12 period $P_{\rm obs} \simeq 6.2\,{\rm ms}$. If PSR B1257+12 has only spun-down since its presumptive initial rapid accretion event, then $P_0$ must be $\lesssim P_{\rm obs}$, constraining the mass-loss exponent to $p \lesssim 0.4$.} \label{fig:SpinPeriod}
\end{figure}

\citet{Miller&Hamilton01} argue that the millisecond pulsar PSR B1257+12 cannot have been recycled by accretion from the same disk responsible for forming the planets, due to the detrimental effects of the large and sustained X-ray accretion luminosity which is associated with the spin-up process.  In the case of a WD-NS merger, however, most of the angular momentum is deposited on the NS during the earliest phases (first $\lesssim$ minute) following the disruption, when the Alfv\'en radius is pushed to the NS surface.
This accretion is capable of spinning up the pulsar to periods of
\begin{equation} \label{eq:P_spin}
P_0 \approx \frac{4p+1}{3} \frac{2 \pi I_{\rm NS}}{M_{\rm d,0} \sqrt{G M_{\rm NS} R_{\rm NS}}} \left(\frac{R_{\rm NS}}{R_{\rm d,0}}\right)^p ,
\end{equation}
as illustrated in Fig~\ref{fig:SpinPeriod}. Here $I_{\rm NS} \simeq 10^{45}\,{\rm g \, cm}^2$, $R_{\rm NS} \simeq 12 \,{\rm km}$ are the NS moment of inertia and radius \citep{Lattimer&Schutz05}. Clearly, rapid post-merger accretion can spin-up the NS to millisecond periods.
At the same time, the disk maintains a reservoir of mass large enough to form planets on much longer timescales, over which the additional accreted mass does not appreciably 
change the pulsar spin. 
Requiring that PSR B1257+12 is spun-up post-accretion to periods shorter than the currently observed $\simeq 6.2\,{\rm ms}$ yields an additional constraint on the mass-loss exponent, $p \lesssim 0.4$.

We additionally speculate that the low X-ray efficiency of PSR B1257+12 (\citealt{Pavlov+07,Yan+13}) may be an artifact of a dramatic rapid-accretion event, since magnetic fields would be buried by such an event, possibly causing the final field topology to differ from that of standard pulsars.

The typical column densities $\sim 10^{3}\,{\rm g\, cm}^{-2}$ of gaseous matter (presumably CO) at onset of solid condensation would shield the disk midplane from impinging X-ray or particle radiation from the NS. Combined with the comparatively short formation timescale implied by the Goldreich-Ward mechanism envisioned in our scenario,
it is reasonable that ablation/evaporation of solids can be averted \citep{Miller&Hamilton01}. 
At late times after disk dispersal, the planets would remain largely unperturbed by the pulsar's hazardous radiation due to their implied diamond composition (boiling temperature $\gtrsim 4000\,{\rm K}$; \citealt{Kuchner&Seager05}).

One of the mysteries of the PSR B1257+12 planetary system is the relatively narrow range of semi-major axes ($\approx 0.2-0.5$ AU) and low eccentricities, which N-body calculations of the final rocky planet assembly by \citet{Hansen+09} indicate requires that the original solid condensation be concentrated over a similarly limited radial range.  The observed outer truncation of the planetary system would appear to favor a situation in which the condensation temperature is achieved immediately once the disk first reaches $R_{\rm p} \sim 0.4$ AU.  The inner truncation could also be explained if solid formation is so efficient that no condensible matter (i.e.~free carbon not locked up in CO) remains in the disk after this initial formation epoch, once the inner regions of the disk cool to the required temperatures.  It is thus a nontrivial feature of our model that values for the solid mass at the $R_{\rm p}$ achieve values close to those required for reasonable values of the disk viscosity $\alpha \sim 10^{-2}$ (Fig.~\ref{fig:ParameterSpace}).

Planets formed from purely carbonaceous grains, as in our scenario, will likely have very different properties, such as mean densities, from those of silicate-dominated planets in our own solar system (e.g., \citealt{Kuchner&Seager05,Mashian&Loeb16}).  Unfortunately, it is challenging to devise of an observation which could confirm the predicted `diamond' composition of the pulsar planets.
Spectroscopy of the planetary atmosphere (if one exists) or surrounding nebular matter seems unfeasible, but could in principle provide such a test.  We would expect a high-metallicity environment dominated by CO and an anomalously large $^{12}$C/$^{13}$C isotopic ratio\footnote{WD carbon matter is expected to be composed of nearly pure $^{12}$C, since any $^{13}$C traces in the WD progenitor's interior would be consumed by rapid $\alpha$-captures during the He burning phase.} $\gg 100$. The latter might distinguish WD-NS merger from SN fallback scenarios, which both predict a high metallicity formation scenario.

The merger of WD-NS binaries have also been proposed to explain the so-called Calcium-rich transients (\citealt{Perets+10,Kasliwal+12}).  These are a class of dim SN-like optical transients which are observed to occur often in the very outskirts of their host galaxies, where very few stars reside.  One of the motivations for associating WD-NS mergers with these events are the birth kicks received by the NS during the SN (\citealt{Metzger12,Lyman+16,Bobrick+16}), which could cause the gravitational wave-driven merger to take place well outside of the galactic plane of the host.  It is thus worth noting that the transverse velocity of PSR B1257+12 is 326 km s$^{-1}$, making it the only known millisecond pulsar with a velocity clearly exceeding $300\,{\rm km \, s}^{-1}$ (\citealt{Yan+13}).  If a connection between WD-NS mergers and the PSR B1257+12 system is confirmed, this in turn links some WD-NS merger systems to the high proper motions necessary to explain the remote positions of the Ca-rich transients.  Though uncertain, the rate of Ca-rich transients is indeed also about 1\% of the core collapse SN rate.  However, the types of binary WDs (He or ONe) giving rise to these explosions may well be different than the C/O WDs which appear to be the most conducive to planet formation.

\section*{Acknowledgments}

We thank Daniel Kasen and Josiah Schwab for helpful information on the opacity of C/O-dominated compositions 
and Cole Miller for helpful comments and suggestions.
The authors gratefully acknowledge support from NASA grants NNX15AU77G (Fermi), NNX15AR47G, NNX16AB30G (Swift), and NNX16AB30G (ATP), NSF grant AST-1410950, and the Alfred P. Sloan Foundation.

\bibliography{PulsarPlanets_Bibliography}

\begin{thebibliography}{}

\bibitem[\protect\citeauthoryear{{Adams}, {Hollenbach}, {Laughlin} \&
  {Gorti}}{{Adams} et~al.}{2004}]{Adams+04}
{Adams} F.~C.,  {Hollenbach} D.,  {Laughlin} G.,    {Gorti} U.,  2004, \apj,
  611, 360

\bibitem[\protect\citeauthoryear{{Armitage}}{{Armitage}}{2010}]{Armitage10}
{Armitage} P.~J.,  2010, {Astrophysics of Planet Formation}

\bibitem[\protect\citeauthoryear{{Balbus} \& {Hawley}}{{Balbus} \&
  {Hawley}}{1991}]{Balbus&Hawley91}
{Balbus} S.~A.,  {Hawley} J.~F.,  1991, \apj, 376, 214

\bibitem[\protect\citeauthoryear{{Blandford} \& {Begelman}}{{Blandford} \&
  {Begelman}}{1999}]{Blandford&Begelman99}
{Blandford} R.~D.,  {Begelman} M.~C.,  1999, \mnras, 303, L1

\bibitem[\protect\citeauthoryear{{Bobrick}, {Davies} \& {Curch}}{{Bobrick}
  et~al.}{2016}]{Bobrick+16}
{Bobrick} A.,  {Davies} M.~B.,    {Curch} R.~P.,  2016, submitted to MNRAS

\bibitem[\protect\citeauthoryear{{Cannizzo}, {Lee} \& {Goodman}}{{Cannizzo}
  et~al.}{1990}]{Cannizzo+90}
{Cannizzo} J.~K.,  {Lee} H.~M.,    {Goodman} J.,  1990, \apj, 351, 38

\bibitem[\protect\citeauthoryear{{Cannizzo}, {Shafter} \& {Wheeler}}{{Cannizzo}
  et~al.}{1988}]{Cannizzo+88}
{Cannizzo} J.~K.,  {Shafter} A.~W.,    {Wheeler} J.~C.,  1988, \apj, 333, 227

\bibitem[\protect\citeauthoryear{{Chevalier}}{{Chevalier}}{1989}]{Chevalier89}
{Chevalier} R.~A.,  1989, \apj, 346, 847

\bibitem[\protect\citeauthoryear{{Chiang} \& {Goldreich}}{{Chiang} \&
  {Goldreich}}{1997}]{Chiang&Goldreich97}
{Chiang} E.~I.,  {Goldreich} P.,  1997, \apj, 490, 368

\bibitem[\protect\citeauthoryear{{Coleman}, {Kotko}, {Blaes}, {Lasota} \&
  {Hirose}}{{Coleman} et~al.}{2016}]{Coleman+16}
{Coleman} M.~S.~B.,  {Kotko} I.,  {Blaes} O.,  {Lasota} J.-P.,    {Hirose} S.,
  2016, \mnras

\bibitem[\protect\citeauthoryear{{Currie} \& {Hansen}}{{Currie} \&
  {Hansen}}{2007}]{Currie&Hansen07}
{Currie} T.,  {Hansen} B.,  2007, \apj, 666, 1232

\bibitem[\protect\citeauthoryear{{Davis}, {Stone} \& {Pessah}}{{Davis}
  et~al.}{2010}]{Davis+10}
{Davis} S.~W.,  {Stone} J.~M.,    {Pessah} M.~E.,  2010, \apj, 713, 52

\bibitem[\protect\citeauthoryear{{Draine} \& {Lee}}{{Draine} \&
  {Lee}}{1984}]{Draine&Lee84}
{Draine} B.~T.,  {Lee} H.~M.,  1984, \apj, 285, 89

\bibitem[\protect\citeauthoryear{{Dullemond} \& {Dominik}}{{Dullemond} \&
  {Dominik}}{2005}]{Dullemond&Dominik05}
{Dullemond} C.~P.,  {Dominik} C.,  2005, \aap, 434, 971

\bibitem[\protect\citeauthoryear{{Fern{\'a}ndez} \& {Metzger}}{{Fern{\'a}ndez}
  \& {Metzger}}{2013}]{Fernandez&Metzger13}
{Fern{\'a}ndez} R.,  {Metzger} B.~D.,  2013, \apj, 763, 108

\bibitem[\protect\citeauthoryear{{Fields}, {Farmer}, {Petermann}, {Iliadis} \&
  {Timmes}}{{Fields} et~al.}{2016}]{Fields+16}
{Fields} C.~E.,  {Farmer} R.,  {Petermann} I.,  {Iliadis} C.,    {Timmes}
  F.~X.,  2016, \apj, 823, 46

\bibitem[\protect\citeauthoryear{{Fleming} \& {Stone}}{{Fleming} \&
  {Stone}}{2003}]{Fleming&Stone03}
{Fleming} T.,  {Stone} J.~M.,  2003, \apj, 585, 908

\bibitem[\protect\citeauthoryear{{Fryer}, {Woosley}, {Herant} \&
  {Davies}}{{Fryer} et~al.}{1999}]{Fryer+99}
{Fryer} C.~L.,  {Woosley} S.~E.,  {Herant} M.,    {Davies} M.~B.,  1999, \apj,
  520, 650

\bibitem[\protect\citeauthoryear{{Gammie}}{{Gammie}}{1996}]{Gammie96}
{Gammie} C.~F.,  1996, \apj, 457, 355

\bibitem[\protect\citeauthoryear{{Goeres}}{{Goeres}}{1996}]{Goeres96}
{Goeres} A.,  1996, in {Jeffery} C.~S.,  {Heber} U.,  eds, Hydrogen Deficient
  Stars Vol.~96 of Astronomical Society of the Pacific Conference Series,
  {Chemistry and thermodynamics of the nucleation in R CrB star shells}.
p.~69

\bibitem[\protect\citeauthoryear{{Goldreich} \& {Ward}}{{Goldreich} \&
  {Ward}}{1973}]{Goldreich&Ward73}
{Goldreich} P.,  {Ward} W.~R.,  1973, \apj, 183, 1051

\bibitem[\protect\citeauthoryear{{Hansen}}{{Hansen}}{2002}]{Hansen02}
{Hansen} B.~M.~S.,  2002, in {Shara} M.~M.,  ed., Stellar Collisions, Mergers
  and their Consequences Vol.~263 of Astronomical Society of the Pacific
  Conference Series, {Stellar Collisions and Pulsar Planets}.
p.~221

\bibitem[\protect\citeauthoryear{{Hansen}, {Shih} \& {Currie}}{{Hansen}
  et~al.}{2009}]{Hansen+09}
{Hansen} B.~M.~S.,  {Shih} H.-Y.,    {Currie} T.,  2009, \apj, 691, 382

\bibitem[\protect\citeauthoryear{{Hawley}, {Balbus} \& {Stone}}{{Hawley}
  et~al.}{2001}]{Hawley+01}
{Hawley} J.~F.,  {Balbus} S.~A.,    {Stone} J.~M.,  2001, \apjl, 554, L49

\bibitem[\protect\citeauthoryear{{Hollenbach}, {Johnstone}, {Lizano} \&
  {Shu}}{{Hollenbach} et~al.}{1994}]{Hollenbach+94}
{Hollenbach} D.,  {Johnstone} D.,  {Lizano} S.,    {Shu} F.,  1994, \apj, 428,
  654

\bibitem[\protect\citeauthoryear{{Iglesias} \& {Rogers}}{{Iglesias} \&
  {Rogers}}{1996}]{Iglesias&Rogers96}
{Iglesias} C.~A.,  {Rogers} F.~J.,  1996, \apj, 464, 943

\bibitem[\protect\citeauthoryear{{Igumenshchev} \& {Abramowicz}}{{Igumenshchev}
  \& {Abramowicz}}{2000}]{Igumenshchev+00}
{Igumenshchev} I.~V.,  {Abramowicz} M.~A.,  2000, \apjs, 130, 463

\bibitem[\protect\citeauthoryear{{Jiang}, {Davis} \& {Stone}}{{Jiang}
  et~al.}{2016}]{Jiang+16}
{Jiang} Y.-F.,  {Davis} S.~W.,    {Stone} J.~M.,  2016, \apj, 827, 10

\bibitem[\protect\citeauthoryear{{Jiang}, {Stone} \& {Davis}}{{Jiang}
  et~al.}{2014}]{Jiang+14}
{Jiang} Y.-F.,  {Stone} J.~M.,    {Davis} S.~W.,  2014, \apj, 796, 106

\bibitem[\protect\citeauthoryear{{Kasliwal} et~al.,}{{Kasliwal}
  et~al.}{2012}]{Kasliwal+12}
{Kasliwal} M.~M.,  et~al., 2012, \apj, 755, 161

\bibitem[\protect\citeauthoryear{{Kerr}, {Johnston}, {Hobbs} \&
  {Shannon}}{{Kerr} et~al.}{2015}]{Kerr+15}
{Kerr} M.,  {Johnston} S.,  {Hobbs} G.,    {Shannon} R.~M.,  2015, \apjl, 809,
  L11

\bibitem[\protect\citeauthoryear{{Kim}, {Perera} \& {McLaughlin}}{{Kim}
  et~al.}{2015}]{Kim+15}
{Kim} C.,  {Perera} B.~B.~P.,    {McLaughlin} M.~A.,  2015, \mnras, 448, 928

\bibitem[\protect\citeauthoryear{{King}, {Olsson} \& {Davies}}{{King}
  et~al.}{2007}]{King+07}
{King} A.,  {Olsson} E.,    {Davies} M.~B.,  2007, \mnras, 374, L34

\bibitem[\protect\citeauthoryear{{Konacki} \& {Wolszczan}}{{Konacki} \&
  {Wolszczan}}{2003}]{Konacki&Wolszczan03}
{Konacki} M.,  {Wolszczan} A.,  2003, \apjl, 591, L147

\bibitem[\protect\citeauthoryear{{Kratter}, {Murray-Clay} \&
  {Youdin}}{{Kratter} et~al.}{2010}]{Kratter+10}
{Kratter} K.~M.,  {Murray-Clay} R.~A.,    {Youdin} A.~N.,  2010, \apj, 710,
  1375

\bibitem[\protect\citeauthoryear{{Kuchner} \& {Seager}}{{Kuchner} \&
  {Seager}}{2005}]{Kuchner&Seager05}
{Kuchner} M.~J.,  {Seager} S.,  2005, ArXiv Astrophysics e-prints

\bibitem[\protect\citeauthoryear{{Laor} \& {Draine}}{{Laor} \&
  {Draine}}{1993}]{Laor&Draine93}
{Laor} A.,  {Draine} B.~T.,  1993, \apj, 402, 441

\bibitem[\protect\citeauthoryear{{Lattimer} \& {Schutz}}{{Lattimer} \&
  {Schutz}}{2005}]{Lattimer&Schutz05}
{Lattimer} J.~M.,  {Schutz} B.~F.,  2005, \apj, 629, 979

\bibitem[\protect\citeauthoryear{{Lazzati} \& {Heger}}{{Lazzati} \&
  {Heger}}{2016}]{Lazzati&Heger16}
{Lazzati} D.,  {Heger} A.,  2016, \apj, 817, 134

\bibitem[\protect\citeauthoryear{{Liebert}, {Bergeron} \& {Holberg}}{{Liebert}
  et~al.}{2005}]{Liebert+05}
{Liebert} J.,  {Bergeron} P.,    {Holberg} J.~B.,  2005, \apjs, 156, 47

\bibitem[\protect\citeauthoryear{{Lightman} \& {Eardley}}{{Lightman} \&
  {Eardley}}{1974}]{Lightman&Eardley74}
{Lightman} A.~P.,  {Eardley} D.~M.,  1974, \apjl, 187, L1

\bibitem[\protect\citeauthoryear{{Luo}, {Brandt}, {Hall}, {Wu}, {Anderson},
  {Garmire}, {Gibson}, {Plotkin}, {Richards}, {Schneider}, {Shemmer} \&
  {Shen}}{{Luo} et~al.}{2015}]{Luo+15}
{Luo} B.,  {Brandt} W.~N.,  {Hall} P.~B.,  {Wu} J.,  {Anderson} S.~F.,
  {Garmire} G.~P.,  {Gibson} R.~R.,  {Plotkin} R.~M.,  {Richards} G.~T.,
  {Schneider} D.~P.,  {Shemmer} O.,    {Shen} Y.,  2015, \apj, 805, 122

\bibitem[\protect\citeauthoryear{{Lyman}, {Levan}, {James}, {Angus}, {Church},
  {Davies} \& {Tanvir}}{{Lyman} et~al.}{2016}]{Lyman+16}
{Lyman} J.~D.,  {Levan} A.~J.,  {James} P.~A.,  {Angus} C.~R.,  {Church} R.~P.,
   {Davies} M.~B.,    {Tanvir} N.~R.,  2016, \mnras, 458, 1768

\bibitem[\protect\citeauthoryear{{MacFadyen} \& {Woosley}}{{MacFadyen} \&
  {Woosley}}{1999}]{MacFadyen&Woosley99}
{MacFadyen} A.~I.,  {Woosley} S.~E.,  1999, \apj, 524, 262

\bibitem[\protect\citeauthoryear{{Margalit} \& {Metzger}}{{Margalit} \&
  {Metzger}}{2016}]{Margalit&Metzger16}
{Margalit} B.,  {Metzger} B.~D.,  2016, ArXiv e-prints

\bibitem[\protect\citeauthoryear{{Marigo} \& {Aringer}}{{Marigo} \&
  {Aringer}}{2009}]{Marigo&Aringer09}
{Marigo} P.,  {Aringer} B.,  2009, \aap, 508, 1539

\bibitem[\protect\citeauthoryear{{Mashian} \& {Loeb}}{{Mashian} \&
  {Loeb}}{2016}]{Mashian&Loeb16}
{Mashian} N.,  {Loeb} A.,  2016, \mnras, 460, 2482

\bibitem[\protect\citeauthoryear{{Mathis}, {Rumpl} \& {Nordsieck}}{{Mathis}
  et~al.}{1977}]{MRN77}
{Mathis} J.~S.,  {Rumpl} W.,    {Nordsieck} K.~H.,  1977, \apj, 217, 425

\bibitem[\protect\citeauthoryear{{McKinney}, {Tchekhovskoy} \&
  {Blandford}}{{McKinney} et~al.}{2012}]{McKinney+12}
{McKinney} J.~C.,  {Tchekhovskoy} A.,    {Blandford} R.~D.,  2012, \mnras, 423,
  3083

\bibitem[\protect\citeauthoryear{{Melis}, {Jura}, {Albert}, {Klein} \&
  {Zuckerman}}{{Melis} et~al.}{2010}]{Melis+10}
{Melis} C.,  {Jura} M.,  {Albert} L.,  {Klein} B.,    {Zuckerman} B.,  2010,
  \apj, 722, 1078

\bibitem[\protect\citeauthoryear{{Menou}, {Perna} \& {Hernquist}}{{Menou}
  et~al.}{2001}]{Menou+01}
{Menou} K.,  {Perna} R.,    {Hernquist} L.,  2001, \apj, 559, 1032

\bibitem[\protect\citeauthoryear{{Metzger}}{{Metzger}}{2012}]{Metzger12}
{Metzger} B.~D.,  2012, \mnras, 419, 827

\bibitem[\protect\citeauthoryear{{Metzger}, {Piro} \& {Quataert}}{{Metzger}
  et~al.}{2008}]{Metzger+08}
{Metzger} B.~D.,  {Piro} A.~L.,    {Quataert} E.,  2008, \mnras, 390, 781

\bibitem[\protect\citeauthoryear{{Miller} \& {Hamilton}}{{Miller} \&
  {Hamilton}}{2001}]{Miller&Hamilton01}
{Miller} M.~C.,  {Hamilton} D.~P.,  2001, \apj, 550, 863

\bibitem[\protect\citeauthoryear{{Narayan}, {S{\"A} dowski}, {Penna} \&
  {Kulkarni}}{{Narayan} et~al.}{2012}]{Narayan+12}
{Narayan} R.,  {S{\"A} dowski} A.,  {Penna} R.~F.,    {Kulkarni} A.~K.,  2012,
  \mnras, 426, 3241

\bibitem[\protect\citeauthoryear{{Nordhaus}, {Spiegel}, {Ibgui}, {Goodman} \&
  {Burrows}}{{Nordhaus} et~al.}{2010}]{Nordhaus+10}
{Nordhaus} J.,  {Spiegel} D.~S.,  {Ibgui} L.,  {Goodman} J.,    {Burrows} A.,
  2010, \mnras, 408, 631

\bibitem[\protect\citeauthoryear{{O'Shaughnessy} \& {Kim}}{{O'Shaughnessy} \&
  {Kim}}{2010}]{OShaughnessy&Kim10}
{O'Shaughnessy} R.,  {Kim} C.,  2010, \apj, 715, 230

\bibitem[\protect\citeauthoryear{{Paschalidis}, {Liu}, {Etienne} \&
  {Shapiro}}{{Paschalidis} et~al.}{2011}]{Paschalidis+11}
{Paschalidis} V.,  {Liu} Y.~T.,  {Etienne} Z.,    {Shapiro} S.~L.,  2011, \prd,
  84, 104032

\bibitem[\protect\citeauthoryear{{Pavlov}, {Kargaltsev}, {Garmire} \&
  {Wolszczan}}{{Pavlov} et~al.}{2007}]{Pavlov+07}
{Pavlov} G.~G.,  {Kargaltsev} O.,  {Garmire} G.~P.,    {Wolszczan} A.,  2007,
  \apj, 664, 1072

\bibitem[\protect\citeauthoryear{{Perets} et~al.,}{{Perets}
  et~al.}{2010}]{Perets+10}
{Perets} H.~B.,  et~al., 2010, \nat, 465, 322

\bibitem[\protect\citeauthoryear{{Perna}, {Duffell}, {Cantiello} \&
  {MacFadyen}}{{Perna} et~al.}{2014}]{Perna+14}
{Perna} R.,  {Duffell} P.,  {Cantiello} M.,    {MacFadyen} A.~I.,  2014, \apj,
  781, 119

\bibitem[\protect\citeauthoryear{{Phinney} \& {Hansen}}{{Phinney} \&
  {Hansen}}{1993}]{Phinney&Hansen93}
{Phinney} E.~S.,  {Hansen} B.~M.~S.,  1993, in {Phillips} J.~A.,  {Thorsett}
  S.~E.,   {Kulkarni} S.~R.,  eds, Planets Around Pulsars Vol.~36 of
  Astronomical Society of the Pacific Conference Series, {The pulsar planet
  production process}.
pp 371--390

\bibitem[\protect\citeauthoryear{{Piran}}{{Piran}}{1978}]{Piran78}
{Piran} T.,  1978, \apj, 221, 652

\bibitem[\protect\citeauthoryear{{Podsiadlowski}}{{Podsiadlowski}}{1993}]{Podsiadlowski93}
{Podsiadlowski} P.,  1993, in {Phillips} J.~A.,  {Thorsett} S.~E.,   {Kulkarni}
  S.~R.,  eds, Planets Around Pulsars Vol.~36 of Astronomical Society of the
  Pacific Conference Series, {Planet formation scenarios}.
pp 149--165

\bibitem[\protect\citeauthoryear{{Pringle}}{{Pringle}}{1981}]{Pringle81}
{Pringle} J.~E.,  1981, \araa, 19, 137

\bibitem[\protect\citeauthoryear{{Pringle}}{{Pringle}}{1991}]{Pringle91}
{Pringle} J.~E.,  1991, \mnras, 248, 754

\bibitem[\protect\citeauthoryear{{Ruderman} \& {Shaham}}{{Ruderman} \&
  {Shaham}}{1985}]{Ruderman&Shaham85}
{Ruderman} M.~A.,  {Shaham} J.,  1985, \apj, 289, 244

\bibitem[\protect\citeauthoryear{{S{\c a}dowski} \& {Narayan}}{{S{\c a}dowski}
  \& {Narayan}}{2015}]{Sadowski&Narayan15}
{S{\c a}dowski} A.,  {Narayan} R.,  2015, \mnras, 453, 3213

\bibitem[\protect\citeauthoryear{{Sekiya}}{{Sekiya}}{1998}]{Sekiya98}
{Sekiya} M.,  1998, \icarus, 133, 298

\bibitem[\protect\citeauthoryear{{Shakura} \& {Sunyaev}}{{Shakura} \&
  {Sunyaev}}{1973}]{Shakura&Sunyaev73}
{Shakura} N.~I.,  {Sunyaev} R.~A.,  1973, \aap, 24, 337

\bibitem[\protect\citeauthoryear{{Shen} \& {Matzner}}{{Shen} \&
  {Matzner}}{2014}]{Shen&Matzner14}
{Shen} R.-F.,  {Matzner} C.~D.,  2014, \apj, 784, 87

\bibitem[\protect\citeauthoryear{{Sirko} \& {Goodman}}{{Sirko} \&
  {Goodman}}{2003}]{Sirko&Goodman03}
{Sirko} E.,  {Goodman} J.,  2003, \mnras, 341, 501

\bibitem[\protect\citeauthoryear{{Stone}, {Pringle} \& {Begelman}}{{Stone}
  et~al.}{1999}]{Stone+99}
{Stone} J.~M.,  {Pringle} J.~E.,    {Begelman} M.~C.,  1999, \mnras, 310, 1002

\bibitem[\protect\citeauthoryear{{van den Heuvel} \& {Bonsema}}{{van den
  Heuvel} \& {Bonsema}}{1984}]{vandenHeuvel&Bonsema84}
{van den Heuvel} E.~P.~J.,  {Bonsema} P.~T.~J.,  1984, \aap, 139, L16

\bibitem[\protect\citeauthoryear{{Verbunt} \& {Rappaport}}{{Verbunt} \&
  {Rappaport}}{1988}]{Verbunt&Rappaport88}
{Verbunt} F.,  {Rappaport} S.,  1988, \apj, 332, 193

\bibitem[\protect\citeauthoryear{{Wang}, {Chakrabarty} \& {Kaplan}}{{Wang}
  et~al.}{2006}]{Wang+06}
{Wang} Z.,  {Chakrabarty} D.,    {Kaplan} D.~L.,  2006, \nat, 440, 772

\bibitem[\protect\citeauthoryear{{Weidenschilling}}{{Weidenschilling}}{1995}]{Weidenschilling95}
{Weidenschilling} S.~J.,  1995, \icarus, 116, 433

\bibitem[\protect\citeauthoryear{{Wolszczan}}{{Wolszczan}}{1994}]{Wolszczan94}
{Wolszczan} A.,  1994, Science, 264, 538

\bibitem[\protect\citeauthoryear{{Wolszczan} \& {Frail}}{{Wolszczan} \&
  {Frail}}{1992}]{Wolszczan&Frail92}
{Wolszczan} A.,  {Frail} D.~A.,  1992, \nat, 355, 145

\bibitem[\protect\citeauthoryear{{Yan}, {Shen}, {Yuan}, {Wang}, {Rottmann} \&
  {Alef}}{{Yan} et~al.}{2013}]{Yan+13}
{Yan} Z.,  {Shen} Z.-Q.,  {Yuan} J.-P.,  {Wang} N.,  {Rottmann} H.,    {Alef}
  W.,  2013, \mnras, 433, 162

\bibitem[\protect\citeauthoryear{{Youdin} \& {Shu}}{{Youdin} \&
  {Shu}}{2002}]{Youdin&Shu02}
{Youdin} A.~N.,  {Shu} F.~H.,  2002, \apj, 580, 494

\bibitem[\protect\citeauthoryear{{Yuan}, {Wu} \& {Bu}}{{Yuan}
  et~al.}{2012}]{Yuan+12}
{Yuan} F.,  {Wu} M.,    {Bu} D.,  2012, \apj, 761, 129

\end{thebibliography}

\appendix
\section{Self-Similar Disk Solutions} \label{sec:Appendix_SelfSimilar}

The set of disk evolution equations presented in \S\ref{sec:DiskModel} are well known to permit a variety of similarity solutions. This Appendix briefly summarizes, extends, and organizes the large number of such solutions presented in the literature \citep[e.g.][and references therein]{Cannizzo+90,Pringle91,Phinney&Hansen93,Metzger+08}. We separately describe the RIAF, radiative, and irradiated phases.

\subsection{RIAF Phase}
During the RIAF phase, the radius and accretion rate of the outer disk evolve in time as
\begin{align}
&R_{\rm d} \propto t^{2/3} ,\\
&\dot{M}_{\rm d} \propto t^{-(2p+4)/3} .
\end{align}
Assuming that radiation provides the dominant component of the midplane pressure, the surface density, temperature and aspect ratio evolve as
\begin{align}
&\Sigma \propto r^{p-1/2} t^{-4(p+1)/3} , \\
&T \propto r^{(p-5/2)/4} t^{-(p+1)/3} , \\
&\theta \propto r^0 t^0 .
\end{align}

\subsection{Radiative Phase}
During the radiative phase, a range of different solutions are permitted, depending on the opacity law and the vertical optical depth of the disk.  Assuming that gas pressure dominates and adopting a general opacity law of the form $\bar{\kappa} \propto \rho^l T^{-k}$ (eq.~\ref{eq:OpacityPowerLaw}), one can manipulate the energy equation to obtain
\begin{equation} \label{eq:theta_m_n}
\theta \propto \Sigma^m r^n
\end{equation} 
where
\begin{equation}
m = \frac{\eta(l+1)+1}{6+\eta(2k+l)} ; ~~~ n = \frac{\eta(k+l)+3/2}{6+\eta(2k+l)},
\end{equation}
\begin{equation}
\eta \equiv \frac{\partial \ln f(\tau)}{\partial \ln \tau} = 
\begin{cases}
1 ; &\tau \geq \sqrt{2/3} \\ 
-1 ; &\tau < \sqrt{2/3},
\end{cases}
\end{equation}
and $f(\tau)$ is our approximation to the flux function given by equation~(\ref{eq:f_of_tau}).

In terms of these shorthand variables, the radius and mass of the disk evolve as
\begin{align} \label{eq:RadiativeRdMdot}
&R_{\rm d} \propto t^{1/(3/2 - 2n + 5m)} , \\
&\dot{M}_{\rm d} \propto t^{-(2 - 2n + 5m)/(3/2 - 2n + 5m)} .
\end{align}
The solution for the local disk variables then follows
\begin{align} \label{eq:RadiativeSigmaTTheta}
&\Sigma \propto r^{-(1/2 + 2n)/(2m + 1)} \dot{M}^{1/(2m + 1)} , \\
&T \propto r^{-(3m - 2n + 1)/(2m + 1)} \dot{M}^{2m/(2m + 1)} , \\
&\theta \propto r^{(n - m/2)/(2m + 1)} \dot{M}^{m/(2m + 1)} .
\end{align}
We have intentionally written the local variables $\Sigma$, $T$, and $\theta$ as a function of the control parameter $\dot{M}$, and do not explicitly substitute equation~(\ref{eq:RadiativeRdMdot}).  This is because the outer disk evolves independently of the inner disk and at most times will reside in an alternative opacity regime, i.e. the values of $m$, $n$ appropriate to the current state of the outer disk can differ from those at some smaller radii where equation (\ref{eq:RadiativeSigmaTTheta}) is to be applied.

\subsection{Irradiated Phase}

Using the same notation as for radiative phase from equation~(\ref{eq:theta_m_n}), we obtain the same class of solutions described by equations~(\ref{eq:RadiativeRdMdot},\ref{eq:RadiativeSigmaTTheta}), except that in this case
\begin{equation}
m = 
\begin{cases}
0 ; &\dot{M} \geq \dot{M}_{\rm Edd} \\
1/5 ; &\dot{M} < \dot{M}_{\rm Edd}
\end{cases}
, ~~~
n = 
\begin{cases}
2/7 ; &\dot{M} \geq \dot{M}_{\rm Edd} \\
1/2 ; &\dot{M} < \dot{M}_{\rm Edd} .
\end{cases}
\end{equation}
The solution in the irradiated regime therefore differs chiefly based on whether the accretion rate is above or below the Eddington limit.

\section{Opacity Curve} \label{sec:Appendix_Opacity}

\begin{figure} 
\centering
\epsfig{file=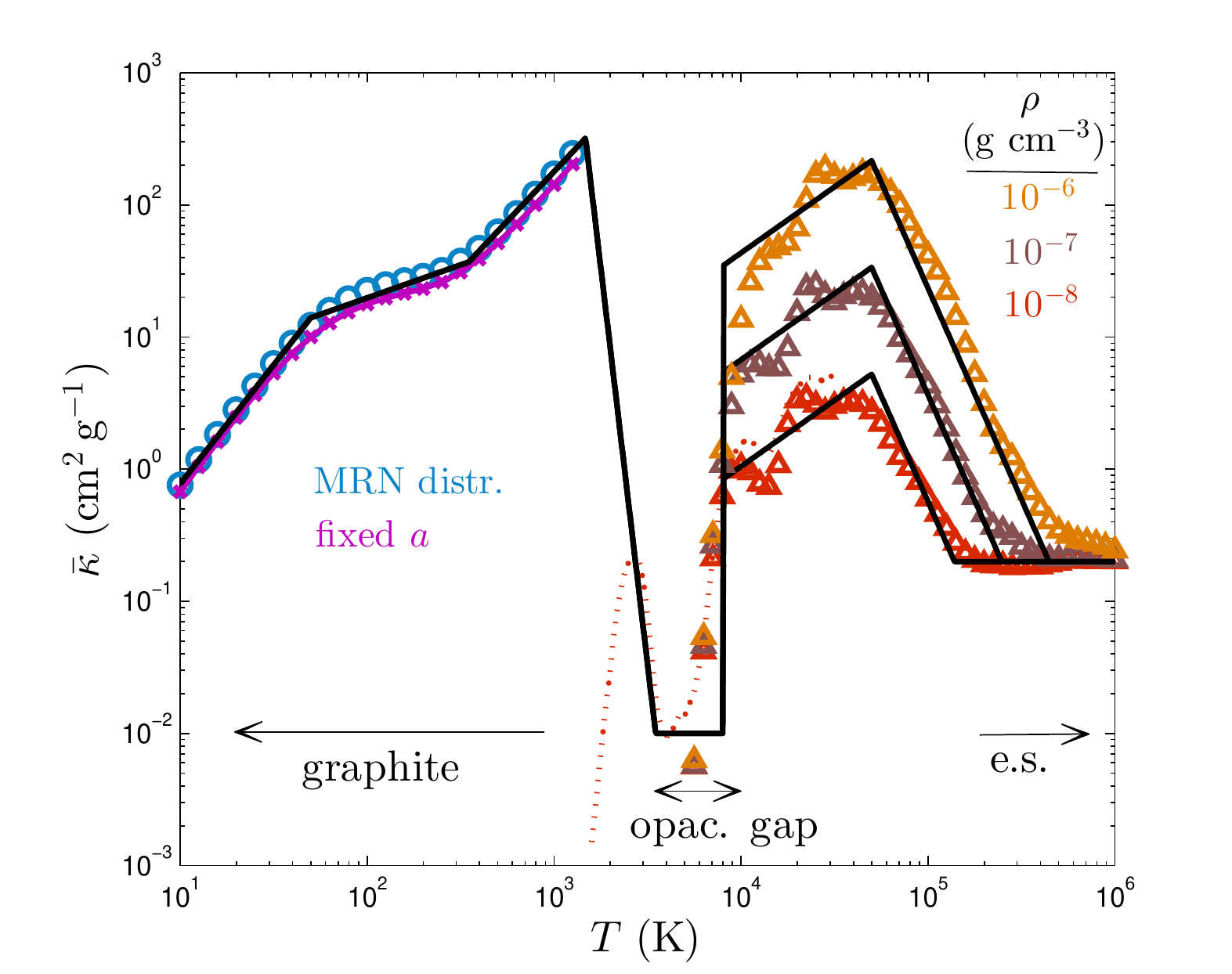,angle=0,width=0.5\textwidth}
\caption{Opacity curve for pure C/O matter as a function of temperature. Open traingles denote opacities obtained from the OPAL project for an ionized carbon-oxygen gas (equal by number). These are plotted at three representative densities (different colors). At lower temperatures, the graphite opacity for an MRN grain size distribution is plotted as open blue circles, and for a fixed grain size of $a = 10^{-3}~\mu {\rm m}$ as purple crosses. The illustrated graphite opacities are for a dust-to-gas ratio of $0.1$, and the results scale linearly with this parameter. At intermediate temperatures, atomic/molecular opacity for $\rho = 10^{-8}\,{\rm g\,cm}^{-3}$ obtained using the AESOPUS code is plotted as the red-dotted curve. The black curves depict our adopted power-law approximation to the opacity curve.} \label{fig:OpacityCurve}
\end{figure}

As shown in Fig.~\ref{fig:OpacityCurve}, we employ a broken power-law approximation to the Planck-averaged mean opacity of the form
\begin{equation} \label{eq:OpacityPowerLaw}
\bar{\kappa} = \bar{\kappa}_0 \rho^l T^{-k},
\end{equation}
which qualitatively mimics the main features expected from the nearly pure carbon/oxygen composition of the disrupted WD disk.  The analytical tractability allowed by this form is convenient for permitting self-similar solutions in the radiative regime (Appendix~\ref{sec:Appendix_SelfSimilar}).  Apart from the recombination ``cliff" below $T \simeq 8000~{\rm K}$, the opacity curve transitions continuously between different regimes.  

At the highest temperatures, where the gas is at least partially ionized, we employ OPAL\footnote{http://opalopacity.llnl.gov/opal.html} opacity tables for an assumed composition of half carbon and half oxygen by mass \citep{Iglesias&Rogers96}.  These are shown as open triangles in Fig.~\ref{fig:OpacityCurve} for three representative densities of $10^{-6}\,{\rm g\,cm}^{-3}$ (orange), $10^{-7}\,{\rm g\,cm}^{-3}$ (brown), and $10^{-8}\,{\rm g\,cm}^{-3}$ (red).  The opacities converge at high temperatures, $T \gtrsim 10^6\,{\rm K}$, where electron scattering dominates ($\bar{\kappa} = 0.2\,{\rm cm}^2\,{\rm g}^{-1}$), as well as at the recombination interface at $T \approx 8000$ K.  At intermediate temperatures, the opacity increases with density, approximately as $\rho^{0.8}$.  Although our parameterization does not capture some of the more subtle details (such as the ``wiggles" around $T \sim 2 \times 10^4\,{\rm K}$), it provides an reasonable first-order approximation to the OPAL opacities.

Below the recombination threshold and before dust condensation, in what is commonly referred to as the `opacity-gap', the opacity is somewhat uncertain. 
Atomic and molecular opacity obtained for our composition using the AESOPUS\footnote{http://stev.oapd.inaf.it/cgi-bin/aesopus} code \citep{Marigo&Aringer09} indicates that the minimal opacity at relevant disk densities 
\begin{align}
\rho_{\rm p} &\approx 3\times 10^{-8}\,{\rm g\,cm}^{-3} \left(\frac{M_{\rm p}}{100M_\oplus}\right) \left(\frac{T_{\rm p}}{2000\,{\rm K}}\right)^{-1/2} \\ \nonumber 
&\times \left(\frac{R_{\rm p}}{0.4\,{\rm AU}}\right)^{-7/2} \left(\frac{\mu}{28}\right)^{1/2}
\end{align}
is $\bar{\kappa} \sim 10^{-2}\,{\rm cm}^{2}\,{\rm g}^{-1}$, and roughly scales as $\sim \rho^{0.5}$.

At temperatures below a few thousand K, solids can condense and grow, and the opacity becomes dominated by dust.  Given the disk composition in our scenario, the dust composition will be dominated by carbonaceous grains (predominantly graphite) because oxygen on its own cannot condense into silicate grains.  We adopt opacities in this range based on the Planck-averaged graphite cross sections calculated by \cite{Draine&Lee84} and \cite{Laor&Draine93}.
These are converted to opacities taking a graphite density of $\simeq 2.2\,{\rm g\,cm}^{-3}$, as shown with open blue circles and purple crosses in Fig.~\ref{fig:OpacityCurve}.  We assume a fiducial dust-to-gas ratio of $10\%$ (see $\S\ref{sec:PlanetFormation}$ for further discussion), and that the graphite opacity simply scales linearly with this ratio if other values are assumed.  The opacity law in this region is well fit by a three-component power-law in temperature (the opacity is essentially independent of density). 

The nominal opacities calculated by \cite{Draine&Lee84} (blue circles in Fig.~\ref{fig:OpacityCurve}) assume a canonical MRN grain size distribution \citep{MRN77} of
\begin{equation}
{\rm d}n \propto a^{-3.5} ~{\rm d}a ~.
\end{equation}
However, a nearly identical opacity law results by instead assuming a single grain size of $a=a_{\rm min} = 10^{-3} ~\mu {\rm m}$ (purple crosses in Fig.~\ref{fig:OpacityCurve}).  This occurs because, for grains smaller than $\sim 0.03~\mu {\rm m}$, the dipole approximation holds well, and the grain cross-section per unit mass is nearly independent of grain size.  Therefore, as long as the grain size distribution decays steeply with $a$ (such as characterizes the MRN distribution), then the opacity will be dominated by grains with $a\sim a_{\rm min}$.  However, as long as $a_{\rm min} \lesssim 0.03~\mu {\rm m}$, then the exact value of $a_{\rm min}$ is inconsequential and hence neither is the precise size distribution.

\end{document}